\documentstyle[12pt,epsfig,times]{article}

\newlength{\dinwidth}
\newlength{\dinmargin}
\begin{document}  
\newcommand{\pom}{{I\!\!P}}
\newcommand{\reg}{{I\!\!R}}
\newcommand{\slowpi}{\pi_{\mathit{slow}}}
\newcommand{\fiidiii}{F_2^{D(3)}}
\newcommand{\fiidiiiarg}{\fiidiii\,(\beta,\,Q^2,\,x)}
\newcommand{\n}{1.19\pm 0.06 (stat.) \pm0.07 (syst.)}
\newcommand{\nz}{1.30\pm 0.08 (stat.)^{+0.08}_{-0.14} (syst.)}
\newcommand{\fiidiiiful}{F_2^{D(4)}\,(\beta,\,Q^2,\,x,\,t)}
\newcommand{\fiipom}{\tilde F_2^D}
\newcommand{\ALPHA}{1.10\pm0.03 (stat.) \pm0.04 (syst.)}
\newcommand{\ALPHAZ}{1.15\pm0.04 (stat.)^{+0.04}_{-0.07} (syst.)}
\newcommand{\fiipomarg}{\fiipom\,(\beta,\,Q^2)}
\newcommand{\pomflux}{f_{\pom / p}}
\newcommand{\nxpom}{1.19\pm 0.06 (stat.) \pm0.07 (syst.)}
\newcommand {\gapprox}
   {\raisebox{-0.7ex}{$\stackrel {\textstyle>}{\sim}$}}
\newcommand {\lapprox}
   {\raisebox{-0.7ex}{$\stackrel {\textstyle<}{\sim}$}}
\def\gsim{\,\lower.25ex\hbox{$\scriptstyle\sim$}\kern-1.30ex%
\raise 0.55ex\hbox{$\scriptstyle >$}\,}
\def\lsim{\,\lower.25ex\hbox{$\scriptstyle\sim$}\kern-1.30ex%
\raise 0.55ex\hbox{$\scriptstyle <$}\,}
\newcommand{\pomfluxarg}{f_{\pom / p}\,(x_\pom)}
\newcommand{\dsf}{\mbox{$F_2^{D(3)}$}}
\newcommand{\dsfva}{\mbox{$F_2^{D(3)}(\beta,Q^2,x_{I\!\!P})$}}
\newcommand{\dsfvb}{\mbox{$F_2^{D(3)}(\beta,Q^2,x)$}}
\newcommand{\dsfpom}{$F_2^{I\!\!P}$}
\newcommand{\gap}{\stackrel{>}{\sim}}
\newcommand{\lap}{\stackrel{<}{\sim}}
\newcommand{\fem}{$F_2^{em}$}
\newcommand{\tsnmp}{$\tilde{\sigma}_{NC}(e^{\mp})$}
\newcommand{\tsnm}{$\tilde{\sigma}_{NC}(e^-)$}
\newcommand{\tsnp}{$\tilde{\sigma}_{NC}(e^+)$}
\newcommand{\st}{$\star$}
\newcommand{\sst}{$\star \star$}
\newcommand{\ssst}{$\star \star \star$}
\newcommand{\sssst}{$\star \star \star \star$}
\newcommand{\tw}{\theta_W}
\newcommand{\sw}{\sin{\theta_W}}
\newcommand{\cw}{\cos{\theta_W}}
\newcommand{\sww}{\sin^2{\theta_W}}
\newcommand{\cww}{\cos^2{\theta_W}}
\newcommand{\trm}{m_{\perp}}
\newcommand{\trp}{p_{\perp}}
\newcommand{\trmm}{m_{\perp}^2}
\newcommand{\trpp}{p_{\perp}^2}
\newcommand{\alp}{\alpha_s}

\newcommand{\alps}{\alpha_s}
\newcommand{\sqrts}{$\sqrt{s}$}
\newcommand{\LO}{$O(\alpha_s^0)$}
\newcommand{\Oa}{$O(\alpha_s)$}
\newcommand{\Oaa}{$O(\alpha_s^2)$}
\newcommand{\PT}{p_{\perp}}
\newcommand{\JPSI}{J/\psi}
\newcommand{\sh}{\hat{s}}
\newcommand{\uh}{\hat{u}}
\newcommand{\MP}{m_{J/\psi}}
\newcommand{\PO}{I\!\!P}
\newcommand{\xbj}{x}
\newcommand{\xpom}{x_{\PO}}
\newcommand{\ttbs}{\char'134}
\newcommand{\xpomlo}{3\times10^{-4}}  
\newcommand{\xpomup}{0.05}  
\newcommand{\dgr}{^\circ}
\newcommand{\pbarnt}{\,\mbox{{\rm pb$^{-1}$}}}
\def\GeV{\hbox{$\;\hbox{\rm GeV}$}}
\newcommand{\WBoson}{\mbox{$W$}}
\newcommand{\fbarn}{\,\mbox{{\rm fb}}}
\newcommand{\fbarnt}{\,\mbox{{\rm fb$^{-1}$}}}
%
%
\newcommand{\qsq}{\ensuremath{Q^2} }
\newcommand{\gevsq}{\ensuremath{\mathrm{GeV}^2} }
\newcommand{\et}{\ensuremath{E_t^*} }
\newcommand{\rap}{\ensuremath{\eta^*} }
\newcommand{\gp}{\ensuremath{\gamma^*}p }
\newcommand{\dsiget}{\ensuremath{{\rm d}\sigma_{ep}/{\rm d}E_t^*} }
\newcommand{\dsigrap}{\ensuremath{{\rm d}\sigma_{ep}/{\rm d}\eta^*} }
\def\Journal#1#2#3#4{{#1} {\bf #2} (#3) #4}
\def\NCA{\em Nuovo Cimento}
\def\NIM{\em Nucl. Instrum. Methods}
\def\NIMA{{\em Nucl. Instr. and Meth.} {\bf A}}
\def\NPB{{\em Nucl. Phys.}   {\bf B}}
\def\PLB{{\em Phys. Lett.}   {\bf B}}
\def\PRL{\em Phys. Rev. Lett.}
\def\PRD{{\em Phys. Rev.}    {\bf D}}
\def\ZPC{{\em Z. Phys.}      {\bf C}}
\def\EJC{{\em Eur. Phys. J.} {\bf C}}
\def\CPC{\em Comp. Phys. Commun.}

\begin{titlepage}

\noindent

\begin{flushleft}
DESY-00-102  \hfill  ISSN0418-9833 \\
July 2000
\end{flushleft}

\vspace{2.2cm}

\begin{center}
\begin{Large}

{\bf A SEARCH FOR EXCITED FERMIONS AT HERA}

\vspace{1.5cm}

H1 Collaboration

\end{Large}
\end{center}

\vspace{1.5cm}

\begin{abstract}
A search for excited fermions ${\tt f^*}$ of the first generation in $e^+p$ scattering at the collider HERA 
is presented using H1 data with an integrated luminosity of 37 pb$^{-1}$.
All electroweak decays of excited fermions, ${\tt f^*} \rightarrow {\tt f} \gamma, {\tt f} W, {\tt f} Z$ 
are considered and all possible final states resulting from the $Z$ or $W$ hadronic decays or decays 
into leptons of the first two generations are taken into account. 
No evidence for ${\tt f^*}$ production is found. Mass dependent exclusion limits on cross-sections and 
on the ratio of coupling constants to the compositeness scale are derived.
\end{abstract}

\vspace{2.5cm}

\begin{center}
To be submitted to The European Physical Journal C
\end{center}

\vfill

\end{titlepage}

%
%
\begin{flushleft}
 C.~Adloff$^{33}$,                
 V.~Andreev$^{24}$,               
 B.~Andrieu$^{27}$,               
 V.~Arkadov$^{35}$,               
 A.~Astvatsatourov$^{35}$,        
 I.~Ayyaz$^{28}$,                 
 A.~Babaev$^{23}$,                
 J.~B\"ahr$^{35}$,                
 P.~Baranov$^{24}$,               
 E.~Barrelet$^{28}$
% ...

\end{flushleft}
%
%
\begin{flushleft} 
  {\it 
 $ ^1$ I. Physikalisches Institut der RWTH, Aachen, Germany$^a$ \\
% ...
$ ^{23}$ Institute for Theoretical and Experimental Physics,
          Moscow, Russia \\
 $ ^{24}$ Lebedev Physical Institute, Moscow, Russia$^{e,h}$ \\
% ...
$ ^{27}$ LPNHE, \'{E}cole Polytechnique, IN2P3-CNRS, Palaiseau, France \\
 $ ^{28}$ LPNHE, Universit\'{e}s Paris VI and VII, IN2P3-CNRS,
          Paris, France \\
% ...
$ ^{33}$ Fachbereich Physik, Bergische Universit\"at Gesamthochschule
          Wuppertal, Wuppertal, Germany$^a$ \\
% ...
$ ^{35}$ DESY, Zeuthen, Germany$^a$ \\
% ...
$ ^a$ Supported by the Bundesministerium f\"ur Bildung, Wissenschaft,
        Forschung und Technologie, FRG,
        under contract numbers 7AC17P, 7AC47P, 7DO55P, 7HH17I, 7HH27P,
        7HD17P, 7HD27P, 7KI17I, 6MP17I and 7WT87P \\
% ...
$ ^e$ Supported by the Deutsche Forschungsgemeinschaft \\
 $ ^f$ Supported by VEGA SR grant no. 2/5167/98 \\
% ...
$ ^h$ Supported by Russian Foundation for Basic Research 
       grant no. 96-02-00019 \\
% ...
 }
\end{flushleft}
\newpage

\boldmath
\section{Introduction}
\unboldmath

Models of composite leptons and quarks~\cite{co1} were introduced in an attempt to provide an 
explanation for the family structure of the known fermions and for their pattern of masses.
A natural consequence of these models is the existence of excited states of leptons and quarks.
It is often assumed that the compositeness scale might be in the ${\rm TeV}$ region, which would 
give excited fermion masses in the same energy domain. However, the dynamics at the constituent 
level being unknown, the lowest excitation states could possibly have masses of the 
order of a few hundred ${\rm GeV}$.
Electron\footnote{The term 'electron' stands generically for electron or positron.}-proton interactions 
at very high energies provide an excellent environment to look for excited fermions of the first generation. 

In this paper a search for excited fermions is presented using $e^+p$ HERA 
collider data of the H1 experiment.
The data collected from 1994 to 1997 at positron and proton beam energies 
of 27.5~${\rm GeV}$ and 820~${\rm GeV}$ respectively correspond to an integrated 
luminosity of 37 pb$^{-1}$.
The excited fermions are searched for through all their electroweak decays into 
a fermion and a gauge boson. 
The subsequent $W$ and $Z$ gauge boson decays considered are those involving 
electrons, muons, neutrinos or jets.
This analysis profits from an increase in statistics by more than a factor of 
10 compared to previous H1 searches~\cite{oldfs1,oldfs3}, and more than a factor of 4 
compared to published results by the ZEUS collaboration~\cite{zeus1}.

The paper is organized as follows. The phenomenological model used to
interpret the results of the search for excited fermions 
is discussed in section~\ref{subsec:pheno}. The H1 detector and the data
preselection criteria are described in section~\ref{subsec:h1d}.
The generators used for the Monte Carlo simulation of the Standard Model events
and excited fermion signals are briefly presented in section~\ref{subsec:gen}. 
The analyses for the various possible final state topologies are described in section~\ref{subsec:ana}. 
The search results are interpreted in section~\ref{subsec:res} and a summary is presented in 
section~\ref{subsec:sum}.

\boldmath
\section{Phenomenology}
\label{subsec:pheno}
\unboldmath

\begin{figure}[hhh]
\begin{center}
\hspace*{-0.4cm}\begin{tabular}{ccc}
 a) & b) & c) \\
\hspace*{-0.25cm}\epsfxsize=0.32\textwidth
 \epsffile{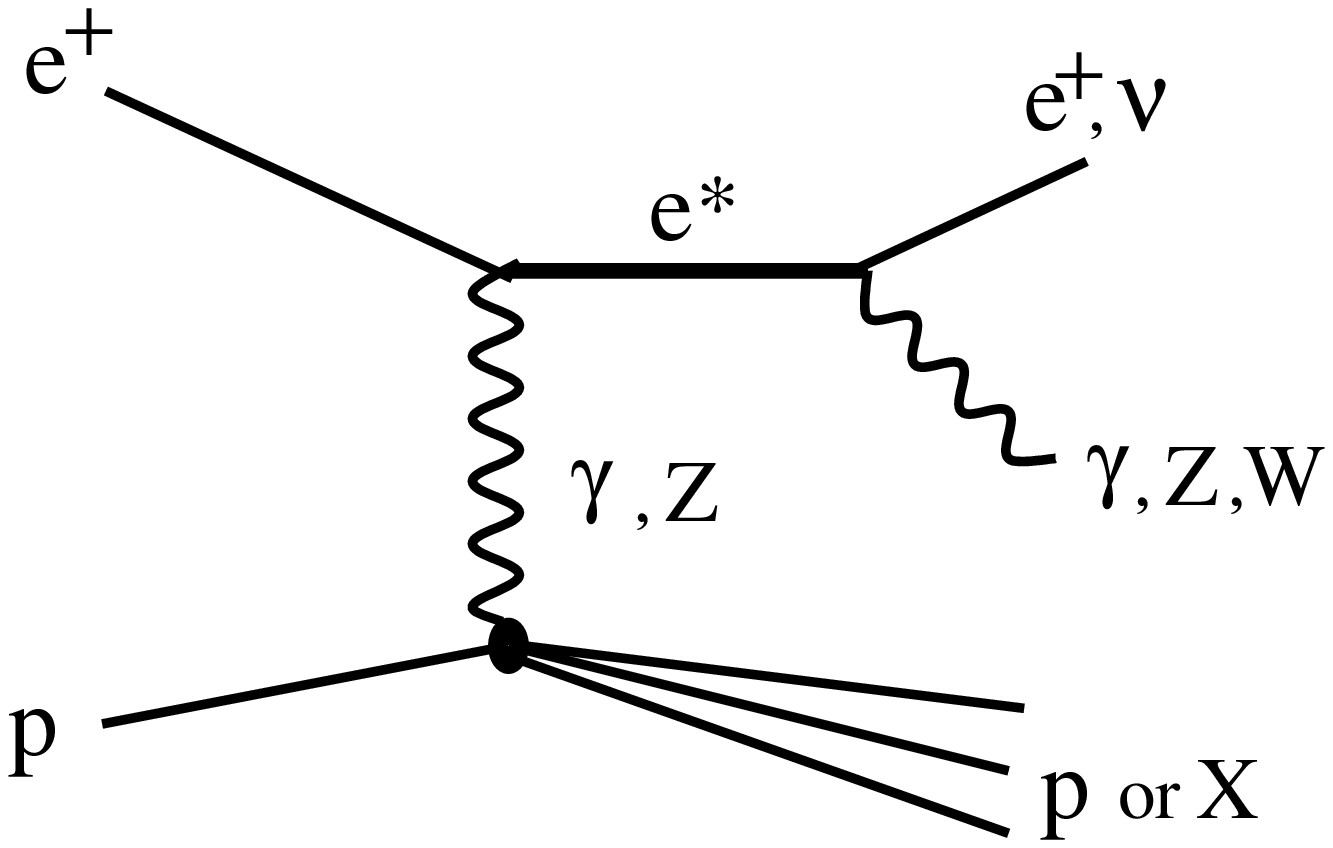} &
\epsfxsize=0.32\textwidth

\hspace*{-0.15cm}\epsffile{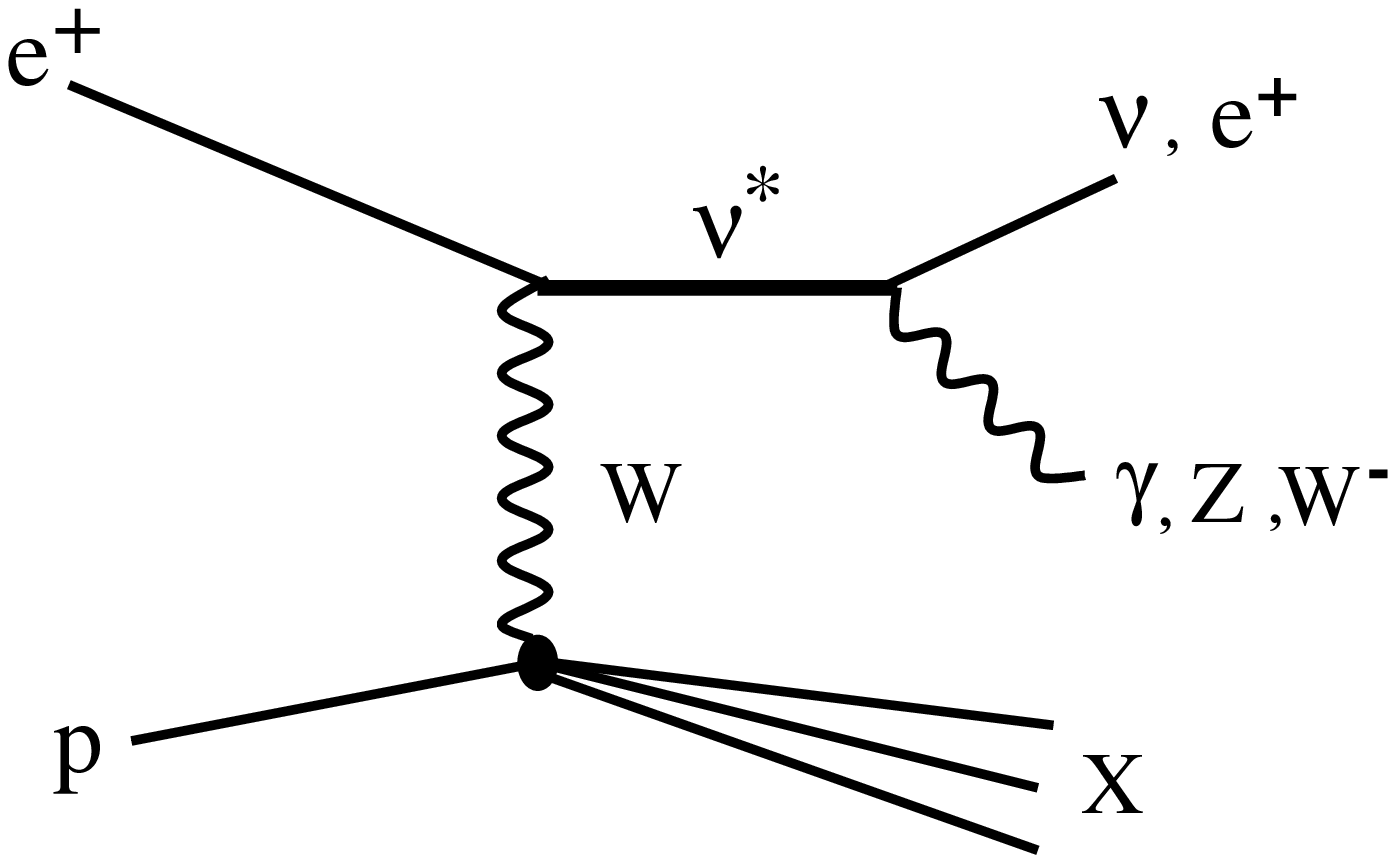}&
\epsfxsize=0.32\textwidth
 \hspace*{-0.45cm}\epsffile{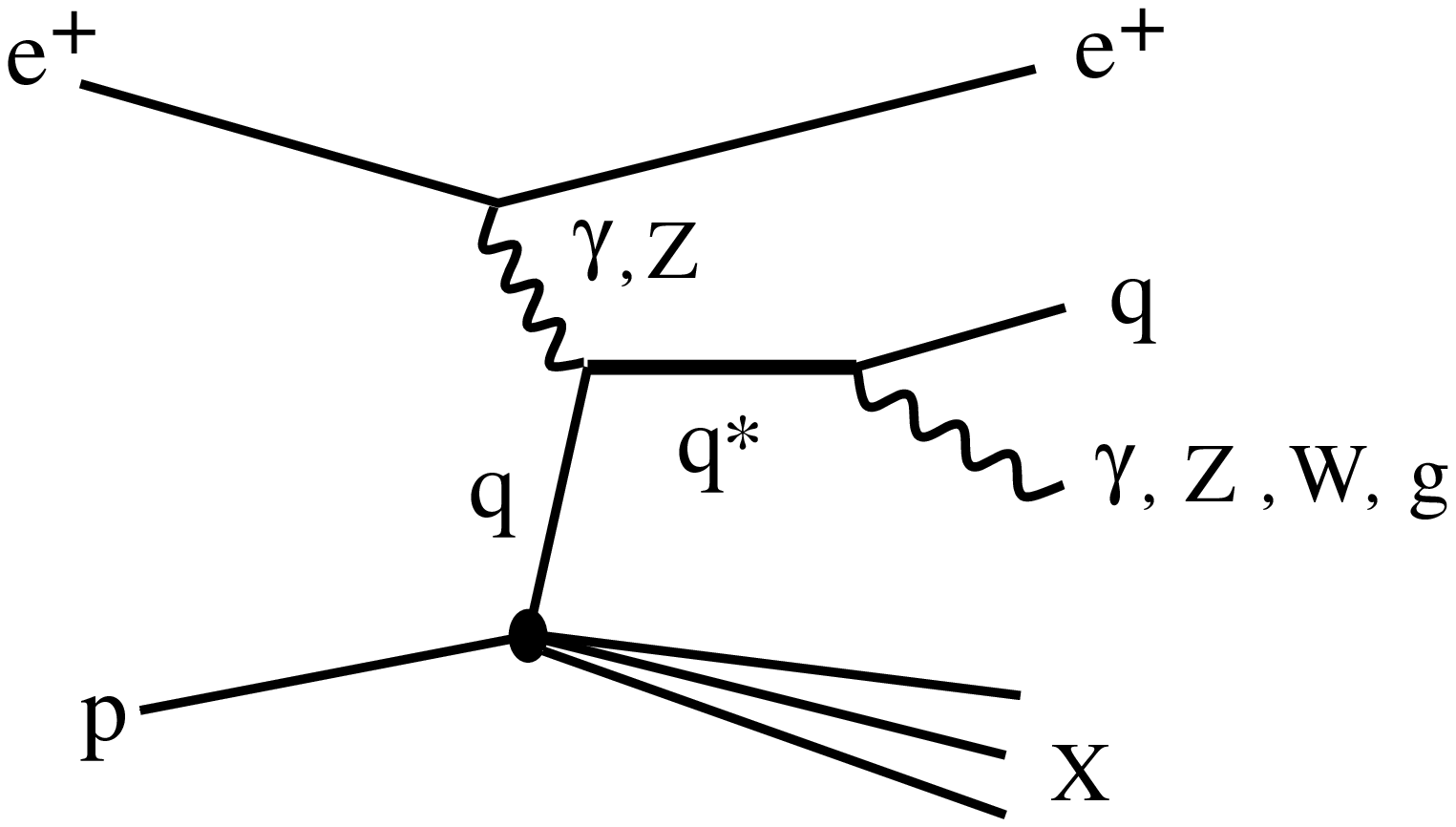}
\\
\end{tabular}
\caption{Leading diagrams for the production and decay of excited fermions in $ep$ collisions.}
\label{fig:graph}
\end{center}
\end{figure}

Excited electrons ($e^*$) could be singly produced in $ep$ collisions through $t-$channel $\gamma$ 
and $Z$ boson exchange~(fig.~\ref{fig:graph}{\it a}). 
Single production of excited neutrinos ($\nu^*$) could result from
$t-$channel $W$ boson exchange~(fig.~\ref{fig:graph}{\it b}).
In the same way excited quarks ($q^*$) could be produced through $t-$channel 
gauge boson exchange between the incoming positron and a quark of 
the proton~(fig.~\ref{fig:graph}{\it c}).

Excited fermions can carry different spin and isospin assignments~\cite{kuhn}. 
In some models quarks and leptons are composites of a scalar and a spin $\frac{1}{2}$ constituent 
and the lowest lying excitations have spin $\frac{1}{2}$.  Alternatively excited
fermions could consist of three spin $\frac{1}{2}$ constituents and in this case the lowest lying 
excitation levels could appear as spin $\frac{3}{2}$ states~\cite{kuhn,jikia}. 
Given that the lowest spin states are often considered as the most probable, we use a 
model~\cite{hagi,baur,boud} in which 
excited fermions are assumed to have spin $\frac{1}{2}$
and isospin $\frac{1}{2}$. This model describes the interaction between the excited fermionic
particles ${\tt f^*}$, the gauge bosons and the ordinary fermionic matter ${\tt f}$ by 
an effective Lagrangian.
Both left-handed and right-handed components of the excited fermions form weak isodoublets $F_L^*$ 
and $F_R^*$. In order to protect the light leptons from radiatively acquiring a large anomalous 
magnetic moment~\cite{brod,rena}, couplings of excited fermions to ordinary fermions of both chiralities 
should be avoided. We choose to consider couplings to left-handed fermions only, in which case 
only the right-handed component of the excited fermions takes part in this magnetic-moment type 
Lagrangian.
The form of the Lagrangian is:
\begin{equation}
 {{\cal{L}}_{int}} = \frac{1}{2\Lambda} {\bar{F^*_R}} \sigma^{\mu \nu} [gf \frac{\tau^a}{2} W_{\mu \nu}^a 
                   + g'f' \frac{Y}{2} B_{\mu \nu} + g_s f_s \frac{\lambda ^a}{2} G_{\mu \nu}^a] {F_L } +  h.c.
\label{lint}
\end{equation}
where $W_{\mu \nu}^a$, $B_{\mu \nu}$ and $G_{\mu \nu}^a$ are the field-strength tensors of the $SU(2)$, $U(1)$
and $SU(3)_C$ gauge fields, $\tau^a$, $Y$ and $\lambda ^a$ are the Pauli matrices, 
the weak hypercharge operator and the Gell-Mann matrices, respectively. 
The $g$, $g'$ and $g_s$ are the standard electroweak and strong gauge couplings. 
$\Lambda$ is the compositeness scale and the $f$, $f'$ and $f_s$ can be viewed as 
form factors (reduced here to parameters) allowing for the composite fermion 
to have arbitrary coupling strengths associated to the three gauge groups.

In this model the coupling constants of the ${\tt f^*}$ with the boson and the fermion ${\tt f}$ are 
related to the parameters $f$ and $f'$ as:
\begin{eqnarray}
c_{\gamma {\tt f^* f}}& = &\frac{1}{2}(f I_3 + f' \frac{Y}{2}) \nonumber \\
c_{Z {\tt f^* f}}& = & \frac{1}{2}(f I_3 \cot \theta _W - f' \frac{Y}{2} \tan  \theta _W)\\
c_{W {\tt f^* f}}& = & \frac{f}{2 \sqrt{2} \sin \theta _W} \nonumber
\label{cconst}
\end{eqnarray}
where $I_3$ is the third component of the isospin of the fermion 
and $\theta _W$ is the Weinberg angle.
The partial widths for the various electroweak decay channels of an excited fermion ${\tt f^*}$
in a fermion ${\tt f}$ and a real boson $V$ are given as~\cite{baur,boud}:
\begin{equation}
\Gamma ({\tt f^*} \rightarrow {\tt f} V) = \alpha \frac{M_{{\tt f^*}} ^3}{\Lambda ^2} c_{V {\tt f^* f}} ^2 
                              (1-\frac{M_{V}^2}{M_{{\tt f^*}}^2})^2 (1+\frac{M_{V}^2}{2M_{{\tt f^*}}^2}) \\
\label{largeur}
\end{equation}
where $M_{{\tt f^*}}$ is the mass of the excited fermion, $M_{V}$ the mass of the 
electroweak boson
and $\alpha$ the fine structure constant.
For the excited quark, the partial width to decay to a quark and a gluon is obtained 
replacing  ($\alpha$) by (4/3 $\alpha _s$) and ($c_{V {\tt f^* f}}$) by (1/2 $f_s$), 
where $\alpha _s$ is the strong coupling constant. 
For $M_{{\tt f^*}}$ values between 50 to 250 ${\rm GeV}$
and $\Lambda = M_{{\tt f^*}}$, the intrinsic widths of the excited electrons are typically 
of the order of some hundred ${\rm MeV}$ for $f$ and $f'$ values $\simeq 1$ while in the 
$q^*$ case for $f_s\simeq 1$ it varies between $1 \GeV$ to $\simeq 10 \GeV$.

\boldmath
\section{The H1 Detector}
\label{subsec:h1d}
\unboldmath

A full description of the  H1 detector can be found in reference~\cite{dete}.
Here we describe briefly the components relevant to this analysis.
The interaction region is surrounded by a system of drift and proportional chambers covering 
the angular range $7^o < \theta < 176^o$. 
The tracking system is placed inside a finely segmented liquid argon (LAr) calorimeter
covering the polar angular range $4^o < \theta < 154^o$.
The electromagnetic part is made of lead/argon and the hadronic part of stainless steel/argon~\cite{lar}. 
Energy resolutions of $\sigma_E / E \simeq 12\% / \sqrt{E} \oplus 1\%$ for electrons and 
$\sigma_E / E \simeq 50\% / \sqrt{E} \oplus 2\%$ for hadrons have been obtained in test beam 
measurements~\cite{testcalo1,testcalo2}.
A lead-scintillating fiber calorimeter\footnote{This detector has replaced in 1995 a conventional 
  lead-scintillator sandwich calorimeter~\cite{bemc}.}~\cite{spacal}
is located in the backward\footnote{The forward direction, $z >$0, 
   from which the polar angle $\theta$ is measured is the proton beam direction.} region
($154^o < \theta < 178^o$) of the H1 detector. 
The tracking system and calorimeters are surrounded by a super-conducting solenoid, producing a uniform 
magnetic field of 1.15 T in the $z$ direction, and an iron yoke instrumented with streamer tubes.
Leakage of hadronic showers outside the calorimeter is measured by analogue charge sampling of the 
streamer tubes with a resolution~\cite{iron} of $\sigma_E / E \simeq 100\% / \sqrt{E}$.
Muon tracks are identified from their hit pattern in the streamer tubes.
   
\boldmath
\section{Event Generators}
\label{subsec:gen}
\unboldmath

Final states of events selected in this analysis contain either a high energy 
electron (or photon) or jets with high transverse energy (or missing transverse momentum).
The main backgrounds from Standard Model processes which could mimic such signatures
are neutral current deep inelastic scattering (NC DIS), charged current 
deep inelastic scattering (CC DIS), photoproduction processes,
QED Compton scattering and $W$ and $Z$ production.

For the determination of the NC DIS contributions we used two 
Monte Carlo samples with different modelling of the QCD radiation: 

\begin{itemize}
\item The first one was produced with the event generator DJANGO~\cite{djan} which includes QED first order 
radiative corrections based on HERACLES~\cite{hera}. 
QCD radiation is implemented using ARIADNE~\cite{aria} based on the Colour Dipole Model (CDM)~\cite{cdm}. 
This sample, with an integrated luminosity of more than 10 times the experimental luminosity, is chosen for 
the estimation of the NC DIS contribution unless explicitly stated otherwise.

\item The second sample was generated with the program RAPGAP~\cite{rapg}, 
where QED first order radiative corrections are implemented as described above. 
RAPGAP includes the leading order QCD matrix element and higher order radiative
corrections are modelled by leading-log parton showers.
This sample of events corresponds to an integrated luminosity of about 2 times the experimental one.
\end{itemize}

For both samples the parton densities in the proton are taken from the MRST~\cite{mrst} parametrization 
which includes constraints from DIS measurements at HERA up to squared momentum transfer 
$Q^2 =$ 5000 ${\rm GeV^2}$~\cite{mrst1,mrst2} and the hadronisation is performed in the Lund
string fragmentation scheme by JETSET~\cite{jets}.

The modelling of the CC DIS process is performed by DJANGO using 
MRST structure functions. QED radiation from quark lines is not fully included 
in the NC DIS and CC DIS simulations.
Whilst inelastic wide angle bremsstrahlung (WAB) is treated in the generator 
DJANGO,
elastic and quasi-elastic WAB (QED Compton scattering) is simulated with the
event generator EPCOMPT~\cite{comp1,comp2}.
Direct and resolved photoproduction processes ($\gamma p$), including prompt 
photon production
are simulated with the PYTHIA~\cite{pyth}
event generator. 
Other processes, corresponding to much lower cross-sections, such as 
lepton pair production or $W$ production have also been simulated. 
The lepton pair production ($\gamma \gamma$) is simulated using the LPAIR generator~\cite{lpair}. 
It should be noted that this generator contains only the Bethe-Heitler $\gamma \gamma$ process. 
However the number of events with two high $E_T$ electromagnetic clusters given by LPAIR and by 
a generator taking into account all (electroweak) tree level graphs and additional 
first order radiative corrections~\cite{dirk} agrees within 5$\%$.
The $W$ simulation is made with the EPVEC program~\cite{epvec}.
The luminosities generated for these Monte Carlo simulations vary between 
3 times to 100 times the experimental luminosity. 

Monte Carlo simulations of excited fermion production and decay
are necessary to evaluate acceptance losses due to selection requirements. 
The excited fermion analyses are based on the phenomenology described in 
section~\ref{subsec:pheno}.
The excited lepton ($l^*$) simulation is performed by the COMPOS~\cite{kohl} 
generator
which makes use of the cross-section formulae given in reference~\cite{hagi}. 
The excited quark generation is done following the cross-section given in 
reference~\cite{jikia}. In both cases initial state radiation of a photon from 
the incoming electron is generated. The photon is taken to be collinear with the
electron, with an energy spectrum given by the Weizs\"{a}cker Williams 
formula. 
The hadronisation is performed here also by the Lund string model and the MRST parametrization 
of the parton densities is used. The narrow-width approximation is assumed and 
the production and decay of 
the excited fermions are assumed to factorize.

All Monte Carlo generators are interfaced
to a full simulation of the H1 detector response.

\boldmath\section{Event Selection and Comparison with Standard Model Expectation}
\label{subsec:ana}\unboldmath

In this section the description of the selection criteria for the analyses 
of the various decay channels is organized according to the experimental 
signatures of the final states. 
Other details of the analyses can be found in reference \cite{isa}. 

In common for all analyses, background not related to $e^+ p$ collisions is 
rejected by requiring that there is a primary vertex within $\pm 35$ cm of 
the nominal vertex value, and that the event time, measured with the central 
tracking chamber, coincides with that of the bunch crossing. 
In addition topological filters against cosmic and halo muons are used. 
A small number of cosmic and halo muons are finally removed by a visual scan. 
 
The identification of electrons or photons, performed in the LAr calorimeter, 
first relies on calorimetric information by exploiting the shape of the energy 
density expected from the development of an electromagnetic 
shower to define electromagnetic clusters.
An electron is identified as an electromagnetic cluster with a track linked to it. 
A photon in contrast should have no track pointing to it within a distance of 40 cm. 
In some analyses electrons and photons are not distinguished and, in this case, 
only electromagnetic (em) clusters are required. 
Hadronic jets (denoted jets in the following) are searched for in the laboratory reference 
frame, using a cone algorithm adapted from the LUCELL scheme from the JETSET package~\cite{jets}, 
with a radius $R = \sqrt {\delta \eta^2 + \delta \phi^2} = 1$, where $\eta$ is the pseudorapidity 
and $\phi$ the azimuthal angle. 
A muon is identified as a well measured central track 
linked geometrically to a track in the muon system or an energy deposit in the instrumented iron. 
A muon candidate should also satisfy an isolation criterium imposed in the pseudorapidity-azimuth 
$(\eta-\phi)$ plane by requiring that the distances of the muon track to the nearest hadronic 
jet and to the closest track be greater than 1 and 0.5 in $\eta$ and $\phi$ respectively.

The event selection makes use of the global variables $(E-P_{z})$, $P^{calo}_{T}$ and 
$P_{T}^{calo \bot}$ described in the following.

\begin{itemize}

\item $(E-P_{z}) =  \sum_{i}(E{_i}-P_{z_{i}}) $
where $E_i$  and $P_{z_{i}} = E{_i} \cos \theta{_i}$ are the energy and the longitudinal momentum 
measured in a calorimeter cell $i$. For an event where the only particles which remain undetected 
are close to the proton direction, momentum conservation implies that $(E-P_{z})$ nearly 
equals twice the energy of the incoming positron (55 ${\rm GeV}$).
 
\item $P^{calo}_{T}$ = $\mid \vec{{P}}_{T}^{calo} \mid$ 
where $\vec{{P}}_{T}^{calo}$ is a missing transverse momentum vector with components calculated
by summing over all energy deposits recorded in cells of the LAr and backward calorimeters.
For the study of the channels including a muon in the final state, this sum is extended to the 
energy deposits in the instrumented iron.
This $P^{calo}_{T}$ variable measures the transverse energy of undetected particles (neutrinos) and 
is sensitive to escaping particles such as high energy muons which leave only a minimum amount of 
energy in the calorimeter. 

\item $P_{T}^{calo \bot}$ defined for events with at least one jet as the 
projection of the vector $\vec{{P}}_{T}^{calo}$ perpendicularly to the jet axis.
For events containing more than one jet the largest among all such projections 
is taken. In channels with missing neutrino signatures, a substantial 
$P_{T}^{calo \bot}$ indicates that the missing momentum is not just due to
fluctuations of the hadronic energy measurement.

\end{itemize}

The selection criteria adapted to the different event topologies are described
in subsections~\ref{subsec:alep} for excited fermions and~\ref{subsec:anaqua} for
excited quarks.
For each of the possible decay channels the number of selected events are compared 
to the Standard Model expectations. The errors given correspond to the statistical and the systematic 
uncertainties added in quadrature. A description of the systematic uncertainties can be found in 
section~\ref{subsec:res}.

\boldmath \subsection{Excited Leptons} \label{subsec:alep}\unboldmath
 
For $e^*$ and $\nu^*$ decays without muons or neutrinos in the final state, 
all particles are detected besides fragments of the proton. For these channels 
a cut 35~$ < (E-P_{z}) <$~65~${\rm GeV}$ is applied to reject photoproduction 
events where one jet is misidentified as an electron or a photon. 
The selection criteria for final states with muons are described in 
subsection~\ref{subsec:mu1}.
For channels involving a neutrino ${P}_{T}^{calo} >$~20~${\rm GeV}$ and 
$ (E-P_{z}) < $~50~${\rm GeV}$ are required, with the additional cut
$P_{T}^{calo \bot} >$~10~${\rm GeV}$ when containing jets. 
For the channels with a $Z$ or $W$ boson in the final state which decays 
via $W \rightarrow q \bar{q}'$, $Z \rightarrow q \bar{q}$ or 
$Z \rightarrow e e$, a reconstructed invariant mass compatible with the
boson mass within $20 \GeV$ is imposed. 
This simple fixed size interval corresponds to about three times 
(seven times) the experimental mass resolution in the case
of hadronic decays (leptonic decays).
A similar mass cut is also imposed in the case of the decay chain
$\nu^*\rightarrow e W \, ; \, W \rightarrow e \nu$ by profiting from
kinematic constraints (subsection~\ref{subsec:ez}).
 
\boldmath 
\subsubsection{The $e^* \rightarrow e \gamma $ channel} 
\label{subsec:eg} 
\unboldmath

The  $e^* \rightarrow e \gamma $ decay mode is the key 
channel to search for $e^*$ because of its very clear signature and
large branching ratio.
The analysis starts from a sample of events with two electromagnetic clusters
in the LAr calorimeter.
The main sources of background 
are the QED Compton process, NC DIS with photon radiation or a high energy
$\pi^0$ in a jet and the two-photon ($\gamma \gamma$) production of
electron pairs. 
Since about half
of the cross-section is expected~\cite{hagi} in the elastic channel 
$e p \rightarrow e^* p$,
the analysis is split into two parts. The first (henceforward called "elastic") 
is dedicated to the search for $e^*$ produced elastically or quasi-elastically, 
the second ("inelastic") concerns the inelastic part 
of the $e^*$ cross-section $e p \rightarrow e^* X$.
 
\begin{itemize}

\item {\bf Elastic channel}
\newline
In this case the signature consists of only two electromagnetic clusters 
and no other signals in the calorimeters. 
The clusters are required to have energies $E_{i} >$ 5 ${\rm GeV}$ and
angles $\theta_i \leq 150^o$ ($i=1,2$), 
with a total energy sum $E_{1} + E_{2} $ above 20 ${\rm GeV}$. 
Because only high invariant masses of excited leptons are of interest, a cut 
$M_{12} > 10 {\rm GeV}$ on the invariant mass $M_{12}$ calculated from the two 
electromagnetic clusters is applied. 
The empty detector condition consists in a cut $E_{tot} - E_{1} - E_{2} < $ 4 ${\rm GeV}$,
where $E_{tot}$ is the total energy deposited in the calorimeters.
This cut strongly suppresses the NC DIS background.
The remaining source of background is the elastic QED Compton scattering. 
\newline
After applying the above cuts
428 events remain. The expectation from Standard Model processes is 
$424 \pm 10$ (418.9 QED Compton and 5.1 $\gamma \gamma$).

\item {\bf Inelastic channel}
\newline
Complementing the elastic analysis, we select here events with 
$E_{tot} -  E_{1} - E_{2} > $ 4 ${\rm GeV}$ and
require two high $E_T$ electromagnetic clusters 
($E_{T_{1}} \geq 20$ and $E_{T_{2}} \geq 10$ ${\rm GeV}$ and $\theta_i \leq 150^o$). 
To reduce the NC DIS contribution with a high energy $\pi^0$ in a jet, a cut
is applied on the multiplicity of tracks ($n_{tracks} \leq 2$) in the direction
close to that of the cluster with the second highest energy.
\newline 
With these criteria $150$ candidates are selected, 
the expected Standard Model background is $ 158 \pm 13 $ events 
(154.7 NC DIS events and 3.3 $\gamma \gamma $). 
\end{itemize}

\begin{table*}[hhh]
  \renewcommand{\doublerulesep}{0.4pt}
  \renewcommand{\arraystretch}{1.2}
 \begin{center}
 \begin{tabular}{p{0.45\textwidth}p{0.55\textwidth}}

    \caption
    {\label{tab:eff1}
    Selection efficiencies in $\%$ for different decay modes of the excited  
    leptons $l^*$ for masses $M_{l^*}$ ranging from 50 to 250 ${\rm GeV}$. 
    The values given for the $e^*\rightarrow e \gamma$ correspond to the 
    combined efficiencies of the elastic and inelastic analyses.}  &
 
\begin{tabular}{|c|c|c|c|c|c|c|}
\hline
$M_{l^*}$ (${\rm GeV}$) & 50 & 75 & 100 & 150 & 200 & 250 \\
\hline
$e^*\rightarrow e \gamma$ & 73 & 85 & 87 & 86 & 83 & 79 \\
$\nu^*\rightarrow \nu \gamma$ & 24 & 41 & 47 & 37 & 32 & 21 \\
\hline
\end{tabular}

\vspace*{0.3cm}

\begin{tabular}{|c|c|c|c|c|c|}
\hline
$M_{l^*}$ (${\rm GeV}$) & 110  & 120 & 150 & 200 & 250 \\
\hline
$e^*\rightarrow e Z_{\rightarrow e e}$ & 71 & 76 & 77 & 77 & 76 \\
$e^*\rightarrow e Z_{\rightarrow \mu \mu }$ & 19 & 43 & 48 & 37 & 28 \\
$e^*\rightarrow e Z_{\rightarrow \nu \bar{\nu}} $ & $<$ 1 & 16 & 72 & 78 & 80 \\
$e^*\rightarrow e Z_{\rightarrow q \bar{q}}$ & 22 & 44 & 46 & 40 & 36 \\ 
$e^*\rightarrow \nu W_{\rightarrow q \bar{q}}$ & 27 & 37 & 44 & 40 & 34 \\
\hline
$\nu^*\rightarrow \nu Z_{\rightarrow e e}$ & 11 & 29 & 40 & 34 & 29 \\
$\nu^*\rightarrow \nu Z_{\rightarrow q \bar{q}}$ & 11 & 25 & 41 & 46 & 37 \\ 
$\nu^*\rightarrow e W_{\rightarrow e \nu }$ & 47 & 45 & 42 & 38 & 32 \\ 
$\nu^*\rightarrow e W_{\rightarrow \mu \nu }$ & 16 & 27 & 40 & 35 & 35 \\
$\nu^*\rightarrow e W_{\rightarrow q \bar{q}}$ & 38 & 39 & 38 & 34 & 28 \\
\hline
 \end{tabular}
  \end{tabular}
  \end{center}
\end{table*}

\begin{figure}[hhh]
\begin{center}
\hspace*{-0.18cm}\begin{tabular}{cc}
 
\epsfxsize=0.52\textwidth
 \epsffile{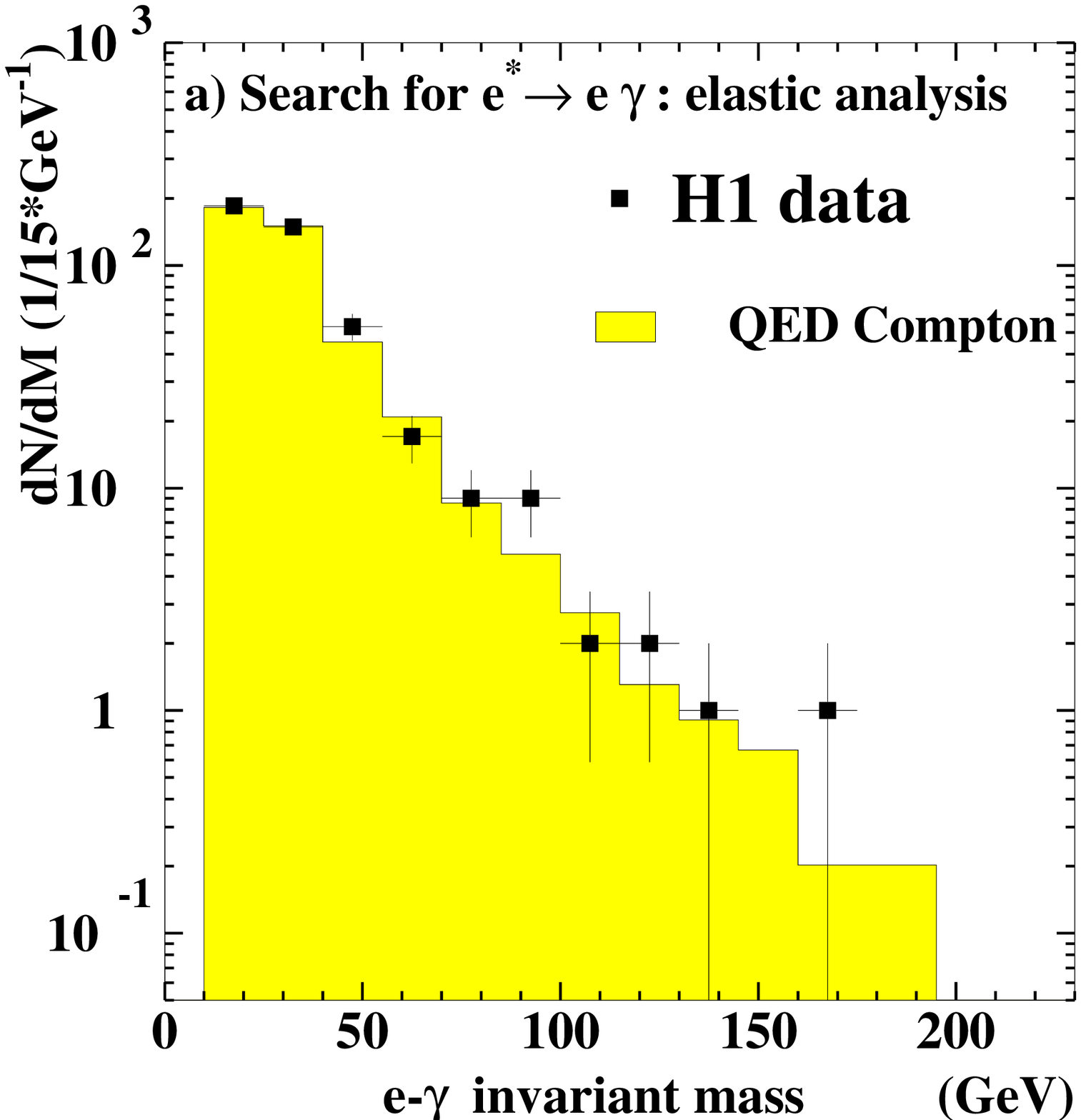} &
\epsfxsize=0.52\textwidth

\hspace*{-0.3cm}\epsffile{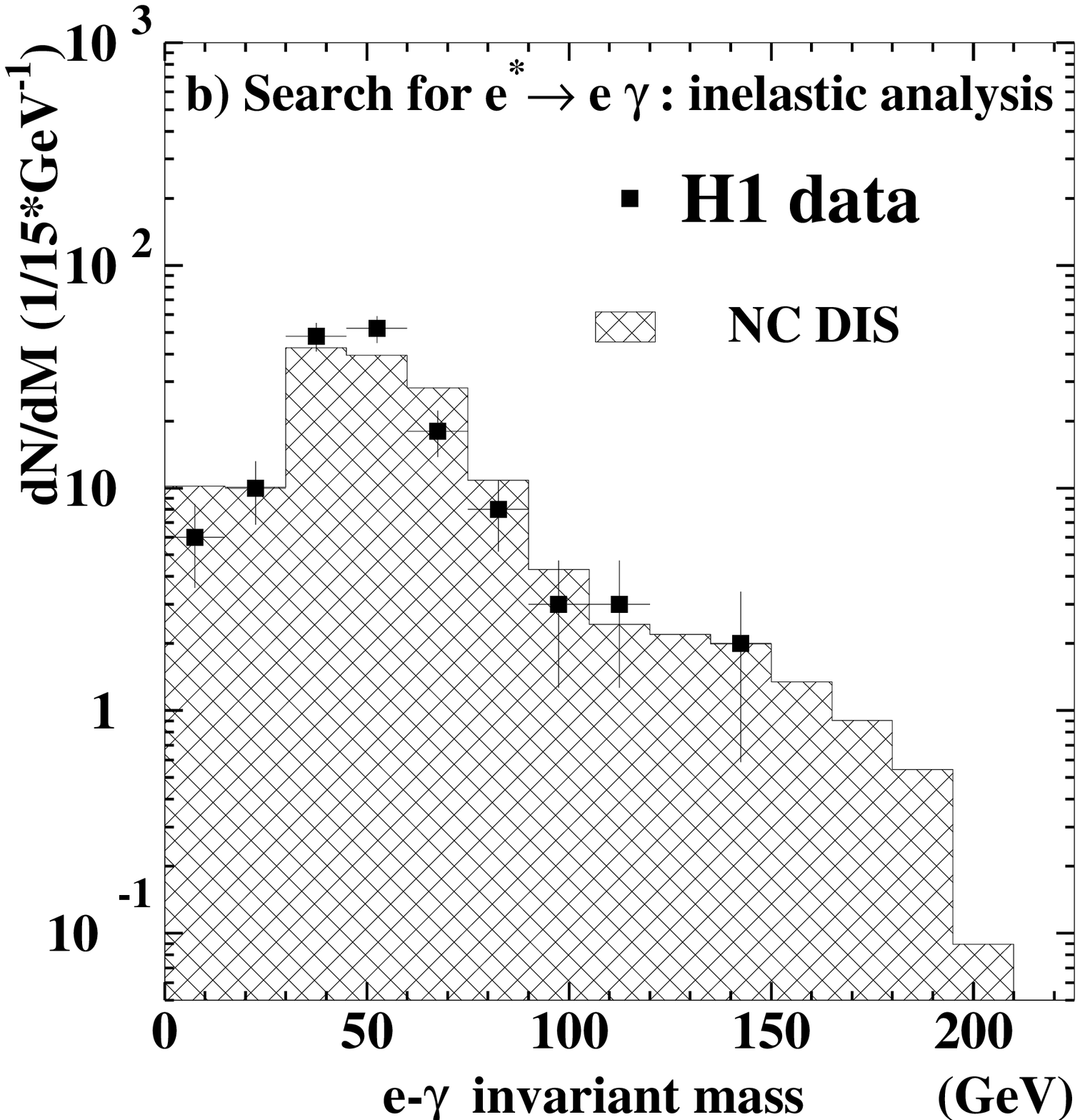} \\

\end{tabular}
	 \caption{Invariant mass spectrum for ({\it a}) the elastic
	 and ({\it b}) the inelastic $e^* \rightarrow e \gamma$
	 analyses.  The invariant mass is calculated for an event by
	 combining the four-momenta reconstructed from the two
	 electromagnetic clusters.  Square symbols correspond to the
	 data and the histograms to the expectations from the
	 different Standard Model processes.}
\label{fig:egmass}
\end{center}
\end{figure}

The combined efficiencies of the two selections are better than $80 \%$
for an $e^*$ mass above 75 ${\rm GeV}$. They are listed 
in table~\ref{tab:eff1}, as well as those of all other $l^*$ decay channels described
later.
The distributions of the measured "$e-\gamma$" invariant mass $M_{12}$ are shown 
together with their corresponding expected background in fig.~\ref{fig:egmass}{\it a} 
and fig.~\ref{fig:egmass}{\it b} for the elastic and inelastic analyses respectively.
The expected mass resolution ($\sigma$ of a Gaussian fit) for the reconstruction of 
an $e^*$ mass of $M_{e^*} =$ 150 ${\rm GeV}$ is 4.2 ${\rm GeV}$.
The number of events with $M_{12}$ above 50~${\rm GeV}$ is 53 in the elastic 
analysis and 68 in the inelastic analysis, in good agreement with the predictions 
from the Standard Model of $52.5 \pm 1.5$ and $77.4 \pm 8.2$ respectively.

\boldmath 
\subsubsection{The $e^* \rightarrow eZ$ and $\nu^* \rightarrow e W, \nu Z$ channels with 
               $Z \rightarrow ee$ and $W \rightarrow e \nu$}
\label{subsec:ez}
\unboldmath

The selection for these channels starts from a sample of events with two high $E_T$ 
electromagnetic clusters ($E_{T_{1}} \geq 20$ and $E_{T_{2}} \geq 10$ ${\rm GeV}$ 
and $\theta_i \leq 150^o$). 

\begin{itemize}
\item {\bf Events with three electromagnetic clusters} 
\newline
In the case of the $e^* \rightarrow e Z_{\rightarrow e e}$  channel,
a third electromagnetic cluster 
with $E_{T_3} \geq $ 5 ${\rm GeV}$ and $\theta_3 \leq 150^o$ is required.
Furthermore at least one pair of the three electromagnetic clusters should have an 
invariant mass compatible with the Z mass. 
To reduce NC DIS background,
events having a jet with $E_{T_{jet}}$ above 15 ${\rm GeV}$ are rejected.
After this selection 1 event is left while
$0.9 \pm 0.4$ background events are expected (0.1 from NC DIS and 0.8 from $\gamma \gamma$ processes).

\item {\bf Events with two electromagnetic clusters and missing \boldmath $P_T$} \unboldmath
\newline
In these cases, the four-momentum of the escaping neutrino is deduced by imposing
transverse momentum conservation and the $(E-Pz)$ constraint.
In the case of the $\nu^* \rightarrow e W_{\rightarrow e \nu} $ and
$\nu^* \rightarrow \nu Z_{\rightarrow e e} $ decay channels, 
events are selected with the invariant mass from the $\nu$ and electromagnetic cluster 
(for $W$ tagging) and from the two electromagnetic clusters (for $Z$ tagging) to be compatible 
with the corresponding boson mass. 
No candidate is found. The expectation from  Standard Model processes is 0.25 $\pm$ 0.11 events 
(0.05 $\gamma \gamma$, 0.1 NC DIS, 0.1 $W \rightarrow e \nu$) for the
$\nu^* \rightarrow e W$ channel, and 0.020 $\pm$ 0.005 $W \rightarrow e \nu$ events for the 
$\nu^* \rightarrow \nu Z$ channel.

\end{itemize}

\boldmath
\subsubsection{The \boldmath $e^* \rightarrow e Z , \nu W $ and $\nu^* \rightarrow \nu Z, e W $ 
               channels with $Z$, $W \rightarrow q \bar{q}'$}
\label{subsec:2je}
\unboldmath
The analysis for these channels uses a subsample of events with at least two jets each having 
a transverse energy greater than 20 ${\rm GeV}$ and a polar angle greater than $10^o$. 
The jet-jet invariant mass must be compatible with the relevant boson mass. 
When more than two jets are found in an event, the pair of jets which has an invariant mass
closest to the relevant boson mass is selected.
This subsample is dominated by photoproduction and NC DIS events.

\begin{itemize}

\item {\bf Events with two high \boldmath $E_T$ jets and one electron} 
\newline
The channels $e^* \rightarrow e Z_{\rightarrow q \bar{q}} $
and $\nu^* \rightarrow e W_{\rightarrow q \bar{q}} $ are characterized by two high $E_T$ jets and an electron. 
Background events are expected from NC DIS.
Candidates are selected if they have an electron with $P_{T_e} \geq 15$ ${\rm GeV}$
and with $ 10^o < \theta_e < 90^o$.
This cut on the transverse momentum of the electron induces an efficiency loss towards low $e^*$ 
masses, already sizeable close to the $Z$ mass (see table \ref{tab:eff1}). 
The cut on the polar angle of the electron discriminates the signal, where the lepton is mainly 
emitted in the forward direction due the high $l^*$ mass, from NC DIS background where the 
electron is mainly scattered in the backward region.
18 events are found in the search for the channel $\nu^* \rightarrow e W$ with an estimated  background of
17.2 $\pm$ 4.8 events (16.6 NC DIS, 0.4 $\gamma p$ and 0.16 $W \rightarrow q \bar{q}'$).
14 events survive the selection criteria for the decay $e^* \rightarrow e Z$, in comparison 
to a background of 12.3 $\pm$ 3.4 events (12 NC DIS, 0.14 $\gamma p$ and 
0.16 $W \rightarrow q \bar{q}'$).
Fig.~\ref{fig:ezmass} shows the invariant masses $M_{ejj}$ of the two jets and the electron.
For a $l^*$ mass of 150 ${\rm GeV}$, the expected mass resolutions are 
8.6 and 13.8 ${\rm GeV}$ for the $e^* \rightarrow e Z$ and 
$\nu^* \rightarrow e W$ channels, respectively. 
The $\nu^*$ mass resolution is worse than the $e^*$ one, because the recoil 
jet is in some cases wrongly taken as one of the jets associated to the $W$ decay. 
No excess of events is found compared to the Standard Model expectation.

\begin{figure}[hhh]
\begin{center}
\hspace*{-0.18cm}\begin{tabular}{cc}
 
\epsfxsize=0.52\textwidth
 \epsffile{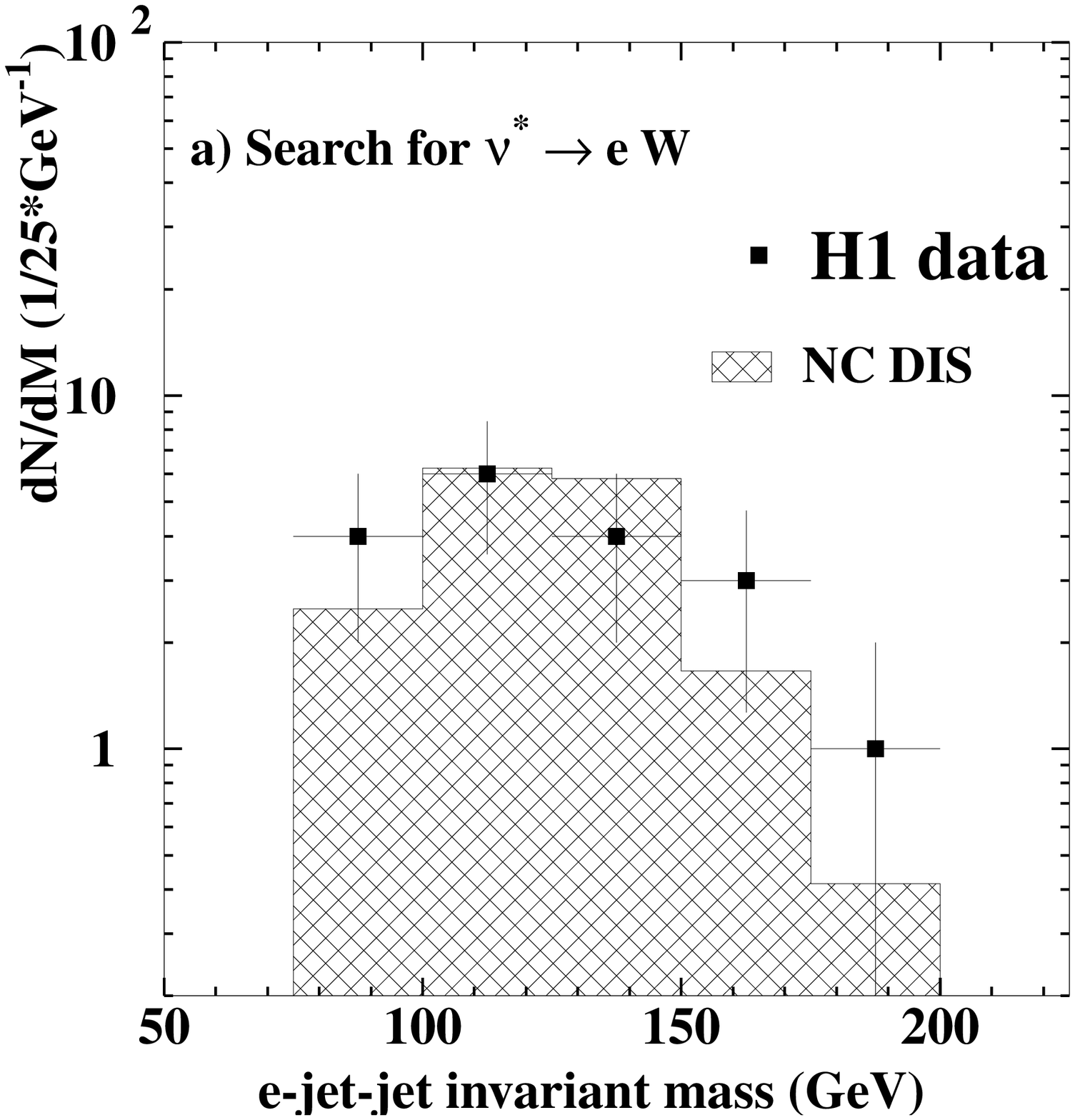} &
\epsfxsize=0.52\textwidth

\hspace*{-0.3cm}\epsffile{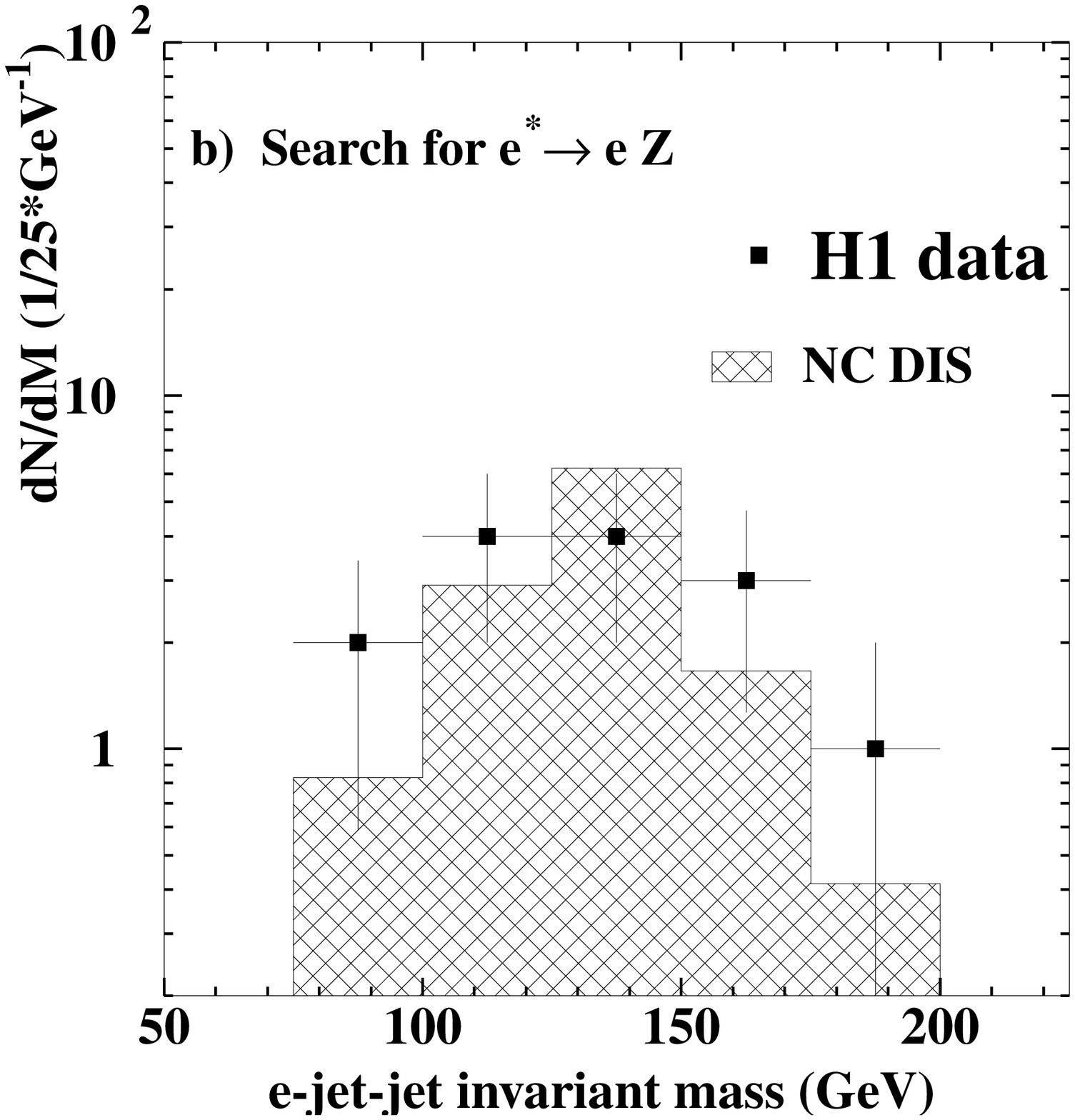} \\
\end{tabular}

\caption{Invariant mass spectrum for the ({\it a}) $\nu^* \rightarrow e W_{\rightarrow q \bar{q}}$ 
         and ({\it b}) $e^* \rightarrow e Z_{\rightarrow q \bar{q}}$ searches.
	 The invariant mass is calculated for an event by combining the four-momenta 
	 reconstructed from the
	 electromagnetic cluster and the two jets. 
	 Square symbols correspond to the data and the histogram to the expectation from 
	 the NC DIS process.}
\label{fig:ezmass}
\end{center}
\end{figure}

\item {\bf Events with two high \boldmath $E_T$ jets and missing $P_T$}
\newline
In the searches $e^* \rightarrow \nu W_{\rightarrow q \bar{q}} $ and
$\nu^* \rightarrow \nu Z_{\rightarrow q \bar{q}} $, the main background which is due to CC DIS 
interactions is suppressed by the ${P}_{T}^{calo}$, $(E-P_{z})$ and $P_{T}^{calo \bot}$ cuts. 
The NC DIS background is reduced by rejecting events possessing an electromagnetic cluster with an energy 
above 5 ${\rm GeV}$.
Three events survive. 
The background expectation is 3.3 $\pm$ 0.6 events (3 CC DIS, 0.3 $\gamma p$)
for the $e^* \rightarrow \nu W_{\rightarrow q \bar{q}} $ channel and
2.1 $\pm$ 0.8 events (2 CC and 0.1 $\gamma p$)
for the $\nu^* \rightarrow \nu Z_{\rightarrow q \bar{q}} $ channel.

\end{itemize}

\boldmath
\subsubsection{The $\nu^* \rightarrow \nu \gamma$ channel}
\label{subsec:etnu}
\unboldmath

For this analysis events containing one photon with $E_{T_{\gamma}} \geq 20$ ${\rm GeV}$ and
$ 10^o < \theta_{\gamma} < 90^o $ and satisfying the ${P}_{T}^{calo}$ and $(E-P_{z})$ cuts 
are selected.
As in subsection~\ref{subsec:2je}, the aim of the $\theta$ cut ($\theta_{\gamma} < 90^o $) is to 
further reduce the NC DIS background for which the electromagnetic 
cluster is predominantly reconstructed in the backward region.
The final state for the signal contains also in most of the cases a recoil jet, 
due to the $\nu^*$ production
through a $t-$channel $W$ boson exchange. 
Hence the final selection criteria are one jet with $E_{T_{jet}} >$ 5 ${\rm GeV}$
and no electron found with an energy above 5 ${\rm GeV}$.
No candidate is left. The expected Standard Model contribution is 
1.0 $\pm$ 0.7 events and is dominated by CC DIS events.

\boldmath
\subsubsection{The $e^* \rightarrow e Z_{\rightarrow \nu \bar{\nu}} $ channel} 
\label{subsec:enu}
\unboldmath

This channel is characterized by one high $P_T$ electron and missing $P_T$ in the detector. 
For this channel the non $ep$ background (cosmic rays and halo muons) is severe and hence the 
minimum $P^{calo}_{T}$ requirement is increased from 20 to 25 ${\rm GeV}$. 
A requirement of large transverse momentum for the
electron is also necessary to reduce the background, so
events with a high $P_T$ electron
($P_{T_e} > $  20 ${\rm GeV}$ with $ 10^o < \theta_e < 150^o$) are selected.
Events with either another electromagnetic cluster of energy greater than 5 ${\rm GeV}$ 
or with a jet with $E_T$ above 15 ${\rm GeV}$ are rejected.
This selection finds 1 event for an expected background of 2.7 $\pm$ 0.4 events 
(1.3 NC DIS, 0.7 $W \rightarrow e \nu$, 0.6 CC DIS and 0.1 $\gamma \gamma$).

\boldmath
\subsubsection{The $\nu^* \rightarrow e W_{\rightarrow \mu \nu}$ and
               $e^* \rightarrow e Z_{\rightarrow \mu \mu}$ channels}
\label{subsec:mu1}
\unboldmath

The search in these channels starts from a subsample of events including at least one
muon candidate found at a polar angle greater then $10^o$ with a transverse momentum 
above 10 ${\rm GeV}$. 

\begin{itemize}

\item {\bf Events with one muon and an electron} 
\newline
For the $\nu^* \rightarrow e W_{\rightarrow \mu \nu}$ analysis we require 
$P^{calo}_{T} \geq $ 25 ${\rm GeV}$ and a high $P_T$ electron ($P_{T_e} > $  20 ${\rm GeV}$).
No events are left after this selection and the total expected background which 
survives these two cuts is 0.31 $\pm$ 0.05 events dominated by 
the muon pair production in $\gamma \gamma$ interactions (0.28 events)
with a small contribution (0.03 events) from single $W \rightarrow \mu \nu$ production.
It has been checked that applying the more stringent cuts of~\cite{muon} 
reduces the $\gamma \gamma$ background by a factor of $\sim$10.

\item {\bf Events with two muons and an electron} 
\newline
The signature of the $e^* \rightarrow e Z_{\rightarrow \mu \mu}$ channel consists of two muons 
plus an electron.
Here, contrary to the preceding one muon case, no neutrino is expected.  
However as the energy deposited in the calorimeter by the two muons is small,
a cut $P^{calo}_{T} \geq$ 15 ${\rm GeV}$ is applied. 
With the requirement of two identified muons no events are left.
The background due to the $\gamma \gamma \rightarrow \mu \mu $ process, 
is 0.35 $\pm$ 0.05 events.

\end{itemize}

\boldmath
\subsection{Excited Quarks}
\label{subsec:anaqua}
\unboldmath

In a $q^*$ production process at HERA one would expect that the scattered positron be often 
unseen in the detector since the process is dominated by $\gamma$ exchange at small values 
of $Q^2$.
Hence no restriction is imposed on the value of $(E-P_{z})$. 
The selection criteria for channels with one or two muons are described in 
subsection~\ref{subsec:mu2}. 
For channels with a neutrino, the cuts $P^{calo}_{T} > $ 20 ${\rm GeV}$
and $P_{T}^{calo \bot} > $ 10 ${\rm GeV}$ are applied.
A compatibility within 20 ${\rm GeV}$ of the Z or W boson mass is imposed here also
in the case of $W \rightarrow q \bar{q}'$, $Z \rightarrow q \bar{q}$ or 
$Z \rightarrow e e$ decays.

\boldmath
\subsubsection{The $q^* \rightarrow q \gamma$ channel} 
\label{subsec:phq}
\unboldmath

\begin{figure}[hhh]
\begin{center}
 \begin{tabular}{p{0.3\textwidth}p{0.7\textwidth}}
    \vspace*{-7cm}
      \caption[]{ \label{fig:qgmass}
          {Invariant mass spectrum for the $q^* \rightarrow q \gamma$ search.
	   The invariant mass is calculated for an event by combining the four-momenta of the 
	   photon and the jet. 
	   Square symbols correspond to the data and the histograms to the 
	   expectations from different Standard Model processes.}}
      &
      \mbox{\epsfxsize=0.55\textwidth
       \epsffile{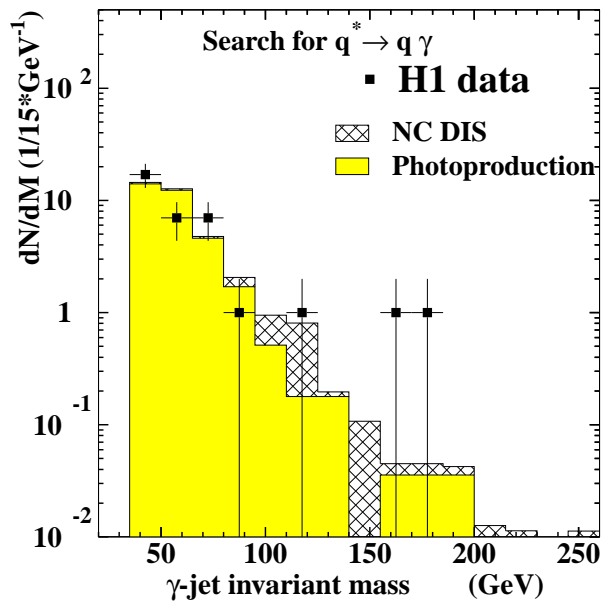}}
\end{tabular}
\end{center}
\end{figure}

The final state for this channel is characterized by one photon and one jet. 
The photon requirements are those described in subsection~\ref{subsec:etnu}: 
one photon with ($E_{T_{\gamma}} \geq 20$ ${\rm GeV}$, 
$ 10^o < \theta_{\gamma} < 90^o $). In addition we require one jet with $E_{T_{jet}}>$ 15 ${\rm GeV}$
and $\theta_{jet} > 10^o$ and no electron with an energy above 5 ${\rm GeV}$ in the LAr calorimeter.
The main background sources are photoproduction processes with prompt $\gamma$ production
or events with high energy $\pi^0$, and NC DIS events if the track of the scattered electron has not 
been reconstructed.
35 events are selected compared to a background estimation of 36 $\pm$ 5 events (2.5 NC DIS and 
33.5  $\gamma p$). 
Fig.~\ref{fig:qgmass} shows the distributions
of the invariant $\gamma$-jet masses for the data and the expected background. 
The expected invariant mass resolution for a $q^*$ mass of 150~${\rm GeV}$ is 6.6~${\rm GeV}$. 
The two events with an invariant mass ($\gamma$-jet) above 150 ${\rm GeV}$ are very likely
NC DIS events, both just surviving the criteria on the quality of the nearest track or on its 
distance to the electromagnetic cluster. 
The efficiency of this selection is listed in table~\ref{tab:eff2}, as well as those of all $q^*$ 
decay channels described later.

\begin{table*}[hhh]
  \renewcommand{\doublerulesep}{0.4pt}
  \renewcommand{\arraystretch}{1.2}
 \begin{center}
 \begin{tabular}{p{0.45\textwidth}p{0.55\textwidth}}

    \caption
    {\label{tab:eff2}
    Selection efficiencies in $\%$ for different decay modes of the excited  
    quarks $q^*$ for masses $M_{q^*}$ ranging from 50 to 250 ${\rm GeV}$.} &

\begin{tabular}{|c|c|c|c|c|c|c|c|}
\hline
$M_{q^*}$ (${\rm GeV}$) & 50 & 75 & 100 & 150 & 200 & 250 \\
\hline
$q^*\rightarrow q \gamma$ & 22 & 33 & 36 & 42 & 41 & 40 \\
\hline
\end{tabular}

\vspace*{0.3cm}

\begin{tabular}{|c|c|c|c|c|c|c|}
\hline
$M_{q^*}$ (${\rm GeV}$) & 110  & 120 & 150 & 200 & 250 \\
\hline
$q^*\rightarrow q Z_{\rightarrow e e} $ & 16 & 35 & 40 & 47 & 41  \\
$q^*\rightarrow q Z_{\rightarrow \mu \mu} $ & 7 & 16 & 25 & 32 & 32  \\
$q^*\rightarrow q Z_{\rightarrow q \bar{q}} $ & $<$ 1  & 3 & 31 & 47 & 44  \\
$q^*\rightarrow q W_{\rightarrow e \nu} $ & 31 & 38 & 42 & 44 & 38 \\
$q^*\rightarrow q W_{\rightarrow \mu \nu} $ & 6 & 24 & 41 & 41 & 37 \\
$q^*\rightarrow q W_{\rightarrow q \bar{q}} $ & 5 & 16 & 42 & 51 & 47 \\
\hline
 \end{tabular}
  \end{tabular}
  \end{center}
\end{table*}

\boldmath 
\subsubsection{The $q^* \rightarrow q Z_{\rightarrow e e}$ channel}
\label{subsec:eez}
\unboldmath

Starting from the subsample of events with at least two electromagnetic clusters described 
in subsection~\ref{subsec:ez}, it is required in addition that there be a jet with 
$E_{T_{jet}}>$ 15 ${\rm GeV}$ and $\theta_{jet} > 10^o$ and that the invariant mass calculated
from the two electromagnetic clusters be compatible with the Z mass.
No events survive these criteria, a fact consistent with the Standard Model expectation of 
$0.65 \pm 0.53$ NC DIS events.

\boldmath
\subsubsection{The $q^* \rightarrow q W_{\rightarrow e \nu}$ channel} 
\label{subsec:etqw}
\unboldmath

For this channel events retained by the ${P}_{T}^{calo}$, $(E-P_{z})$ and $P_{T}^{calo \bot}$ 
cuts and with only one high $P_T$ electron ($P_{T_e} > $ 15 ${\rm GeV}$
and $ 10^o < \theta_{e} < 150^o $) and a jet ($E_{T_{jet}} > $ 15 ${\rm GeV}$ and 
$\theta_{jet} > 10^o $) are selected.
The main sources of background in this channel are NC and CC DIS processes and 
$W$ production. 
One event survives this selection and the number of events expected from the Standard Model is 
1.10 $\pm$ 0.35 events, equally shared between the NC DIS and $W \rightarrow e \nu$ processes.

\boldmath
\subsubsection{The $q^* \rightarrow q W_{\rightarrow q \bar{q}'} $ 
               and $q^* \rightarrow q Z_{\rightarrow q \bar{q}}$  channels} 
\label{subsec:qjj}
\unboldmath

\begin{figure}[hhh]
\begin{center}
\hspace*{-0.18cm}\begin{tabular}{cc}
 
\epsfxsize=0.52\textwidth
 \epsffile{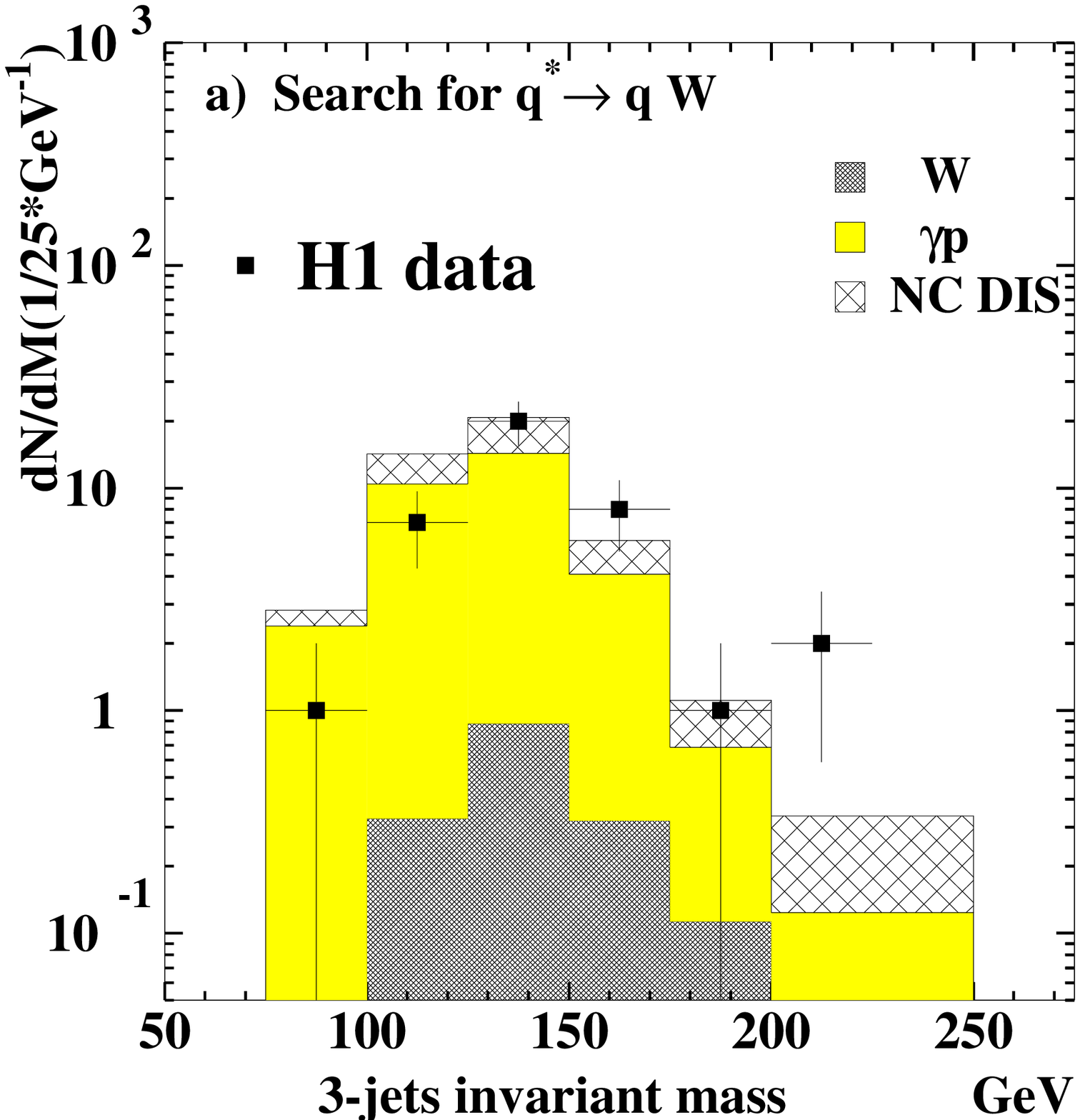} &
\epsfxsize=0.52\textwidth

\hspace*{-0.3cm}\epsffile{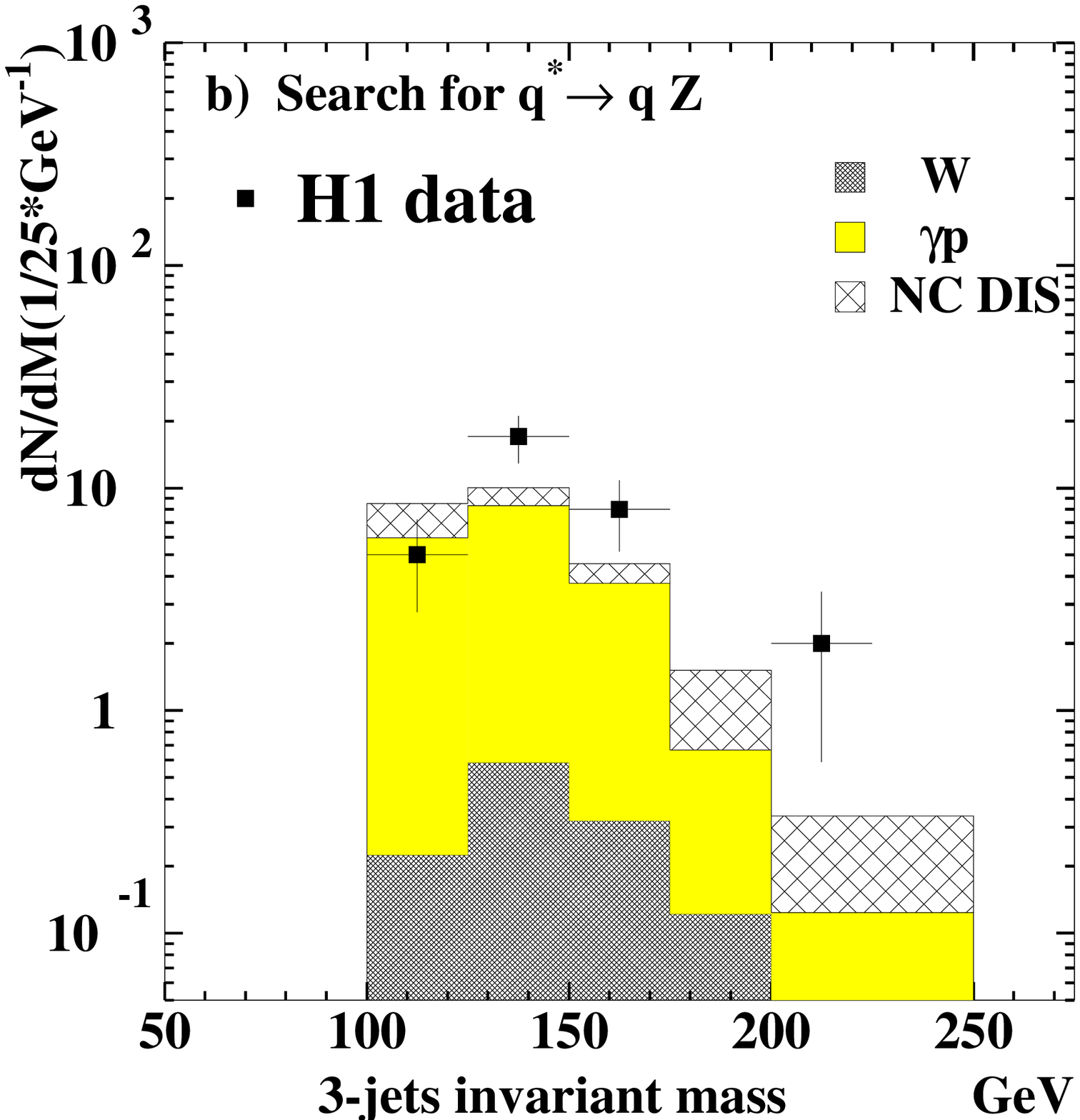} \\
\end{tabular}

\caption{Invariant mass spectrum for ({\it a}) the $q^* \rightarrow q W_{\rightarrow q \bar{q}}$ 
         and ({\it b}) the $q^* \rightarrow q Z_{\rightarrow q \bar{q}} $ analyses.
         The invariant mass is calculated for an event by combining the four-momenta of
	 the three jets. Square symbols correspond to the data and the histograms to the 
         expectations from different Standard Model processes.}
\label{fig:qqqmass}

\end{center}
\end{figure}

The final state in these channels contains three high $E_T$ jets and the main backgrounds are 
photoproduction and NC DIS processes. 
We require three jets with a polar angle above $10^o$ and transverse energies greater than 30, 25 
and 15 ${\rm GeV}$, respectively.
Furthermore, in these $q^*$ decays, the jet with lowest transverse energy often originates
from the boson (W/Z) decay, when the $q^*$ mass is above 150 ${\rm GeV}$.
Events are kept only when the jet-jet invariant mass calculation which is the nearest to 
the W/Z mass includes this jet. 
For the channel $q^* \rightarrow q Z_{\rightarrow q \bar{q}} $, the loss of efficiency due to this 
requirement varies between 30$\%$ for a $q^*$ mass of 150 ${\rm GeV}$ to less than one percent 
when the $q^*$ mass is equal to 250 ${\rm GeV}$. 
In the $q^* \rightarrow q W_{\rightarrow q \bar{q}} $ case the losses are smaller. 
We require that this invariant jet-jet mass be compatible with the boson mass.
These cuts select 39 and 32 events for the $q^* \rightarrow q W_{\rightarrow q \bar{q}} $ and
the $q^* \rightarrow q Z_{\rightarrow q \bar{q}}$ channels, respectively.
These numbers are to be compared to the Standard Model expectations of  45.3 $\pm$ 17.3 events
(30.4 $\gamma p$, 13.2 NC DIS and 1.7 W) for the $q^* \rightarrow q W $ analysis, and of 
25.3 $\pm$ 9.1 events (17.6 $\gamma p$, 6.4 NC DIS and 1.3 W) for $q^* \rightarrow q Z $. 
The NC DIS expectation has been calculated using the generator RAPGAP, in which leading log parton 
showers are used to model QCD radiations. 
The DJANGO generator which uses the CDM to simulate QCD effects is not able to describe the data 
in this particular phase space domain. 
A similar observation in the measurement of 2-jet rates in DIS has been reported 
previously~\cite{hrap}. 
Detailed investigations of discrepancies between different QCD cascade models are underway~\cite{work}.
The shapes of the invariant three-jet mass distributions are in good agreement with the 
Standard Model expectations, as can be seen in fig.~\ref{fig:qqqmass}. For a $q^*$ mass of 
150 ${\rm GeV}$, the expected resolution on the 3-jet invariant mass is 9 ${\rm GeV}$.

\boldmath
\subsubsection{The $q^* \rightarrow q W_{\rightarrow \mu \nu}$ and
               $q^* \rightarrow q Z_{\rightarrow \mu \mu}$ channels}
\label{subsec:mu2}
\unboldmath

These analyses use the muon subsample described in subsection~\ref{subsec:mu1},
together with the cuts $P^{calo}_{T} \geq $ 25 ${\rm GeV}$ and 
$P^{calo}_{T} \geq $ 15 ${\rm GeV}$ 
for the one muon and two muon searches, respectively.
\begin{itemize}

\item {\bf Events with one muon and a jet}
\newline
In the $q^* \rightarrow q W_{\rightarrow \mu \nu}$ search, events including a high $E_T$ jet 
($E_{T_{jet}} > $ 25 ${\rm GeV}$ and $\theta_{jet} > 10^o$) are selected. 
To reduce the $\gamma \gamma \rightarrow \mu \mu $ background, events with more than one isolated 
muon are rejected and an acoplanarity cut $\Delta \phi(\mu-jet) < 175 ^0$ in the transverse plane 
is applied. 
Three events are found. They correspond to the events labelled muon-2,
muon-4 and muon-5 observed already in a previous search for events with high $P_T$ leptons and a 
large missing transverse momentum~\cite{muon}. 
The background (0.41 $\pm$ 0.03 events) is here due to W production (0.31 events) and 
$\gamma \gamma$ muon pair production (0.1 events).

\item {\bf Events with two muons and a jet} 
\newline
The signature of the $q^* \rightarrow q Z_{\rightarrow \mu \mu}$
channel consists of two muons plus a jet.
When requesting two muons and applying a $P^{calo}_{T} > 15 \GeV$ cut no events remain. 
The background is equal to 0.35 $\pm$ 0.05 events from the 
$\gamma \gamma \rightarrow \mu \mu $ process.

\end{itemize}

\boldmath\section{ Limits on Excited Fermion Production}\label{subsec:res}\unboldmath 

\boldmath\subsection{ Upper Limits on Cross-sections}
\label{subsec:sig}\unboldmath 
A summary of the number of events surviving the selection cuts in the various channels is given
in table~\ref{tab:res1} for the excited lepton decay channels and in table~\ref{tab:res2} for the 
excited quark decay channels.
The uncertainties taken into account on the background determination are listed in the following.
\begin{itemize}
\item The statistical error of the Monte Carlo generations.
\item An uncertainty on the absolute electromagnetic energy scale ranging from $\pm 0.7 \%$ 
      in the central part of the LAr calorimeter to $\pm 3 \%$ in the forward part.
\item An uncertainty of $\pm 4 \%$ on the energy of the jets due to the uncertainty on the 
      calibration of the calorimeters for hadronic showers.
\item A $7 \%$ uncertainty on the predicted DIS cross-sections coming mainly from the lack of 
      knowledge on the proton structure (see detailed discussion in~\cite{hipt}).
\item An uncertainty of $\pm 10 \%$ on the expectation for the 2-jet cross-section estimated by 
      comparing leading order and next-to-leading order Monte Carlo simulations. 
      In the same way an uncertainty of $\pm 15 \%$ on the three-jet cross-section was determined 
      by a comparison to data of either a Monte Carlo with $O(\alpha_s)$ QCD matrix elements which 
      approximates the higher order emission of partons using the concept of parton showers, or 
      perturbative QCD calculations to order $O(\alpha^2_s)$ which produces an exact leading order 
      calculation of the three parton final state~\cite{errjets}.
\item An uncertainty on the estimation of the radiative CC DIS background for the 
      $\nu^* \rightarrow \nu \gamma$ channel, coming from the fact that photon radiation from the 
      quark lines is not fully taken into account in our CC DIS simulation. 
      Calculations~\cite{hel92} show that the negative interference term between photon 
      radiation from the electron and quark lines could decrease the radiative CC DIS 
      cross-section by an amount of $70 \%$.
\item An overall systematic error of $1.5\%$ on the luminosity.

\end{itemize}

The observed number of events are compared to the expected Standard Model background in 
tables~\ref{tab:res1} and \ref{tab:res2}. Good agreement is found for all channels, 
except for a slight excess, already quoted in reference~\cite{muon}, 
observed in the $q^* \rightarrow q W_{\rightarrow \mu \nu}$ channel where 3 events are observed
for a Standard Model expectation of 0.41 $\pm$ 0.03 events.
However, combining the contributions from the three $W$ decay channels, 
no significant 
deviation to the Standard Model prediction remains within the present analysis.

\begin{table}[hhh]
\begin{center}

\begin{tabular}{|c|c|c|c|}
\hline
 Channel & selection criteria & events  & background \\
\hline
$e^*\rightarrow e \gamma$ (el. an.) & 2 em clusters only & 53 & 52.5 $\pm$ 1.5 \\
$e^*\rightarrow e \gamma$ (inel. an.) & 2 high $E_T$ em clusters & 68 & 77.4 $\pm$ 8.2 \\
$e^*\rightarrow e Z_{\rightarrow e e}$ & 3 em clusters & 1 & 0.9 $\pm$ 0.4\\
$e^*\rightarrow e Z_{\rightarrow \mu \mu }$ & 2 muons + 1 electron & 0 & 0.35 $\pm$ 0.05 \\
$e^*\rightarrow e Z_{\rightarrow \nu \bar{\nu}} $ & 1 electron + $P_{T}^{miss}$ & 1 & 2.7 $\pm$ 0.4 \\
$e^*\rightarrow e Z_{\rightarrow q \bar{q}}$ & 2 jets + 1 electron & 14 & 12.3 $\pm$ 3.4 \\ 
$e^*\rightarrow \nu W_{\rightarrow q \bar{q}}$ & 2 jets + $P_{T}^{miss}$ & 3 & 3.3 $\pm$ 0.6 \\
\hline
$\nu^*\rightarrow \nu \gamma$ & 1 photon + $P_{T}^{miss}$ & 0 &  1.0 $\pm$ 0.7 \\
$\nu^*\rightarrow \nu Z_{\rightarrow e e}$ & 2 em clusters + $P_{T}^{miss}$ & 0 & 0.020 $\pm$ 0.005\\
$\nu^*\rightarrow \nu Z_{\rightarrow q \bar{q}}$ & 2 jets + $P_{T}^{miss}$ & 3 & 2.1 $\pm$ 0.8 \\ 
$\nu^*\rightarrow e W_{\rightarrow e \nu }$ & 2 em clusters + $P_{T}^{miss}$ & 0 & 0.25 $\pm$ 0.11 \\ 
$\nu^*\rightarrow e W_{\rightarrow \mu \nu }$ & 1 muon + 1 electron & 0 & 0.31 $\pm$ 0.05 \\
$\nu^*\rightarrow e W_{\rightarrow q \bar{q}}$ & 2 jets + 1 electron & 18 & 17.2 $\pm$ 4.8 \\
\hline
\end{tabular}
\caption{Number of events observed in the various $e^*$ and $\nu^*$ decay channels and the 
         corresponding Standard Model expectation and total uncertainty on the mean expectation.
         It should be noted that these numbers correspond to different invariant mass intervals, 
	 as the effective mass threshold depends on the channel.}
\label{tab:res1}

\end{center}
\end{table}

No evidence was seen for either excited leptons or quarks in any of the channels.
Therefore, upper limits on the product of the ${\tt f^*}$ production cross-section and the decay
branching fraction have been derived.
These limits are determined at a Confidence Level (CL) of $95 \%$ as a function of the excited 
fermion mass. 
A mass window is shifted over the whole mass range in steps of 5 ${\rm GeV}$. 
The width of each window is chosen according to the resolution for the corresponding mass. 
When combining several decay channels, for each decay channel $k$ and in each mass interval, the 
number of observed events $n_k$, the number of expected background events $b_k$ and $\epsilon_k$, 
the product of the efficiency times branching ratio of the channel, are calculated and used
to determine the value A of the upper limit for the signal such that:
\begin{equation}
   CL = \int_{0}^{A} p(a) da / \int_{0}^{\infty} p(a) da \,\,\, ; \,\,\,
         p(a) = \prod_k \frac{1}{n_k!} (\epsilon_k a+b_k)^{n_k}  \exp^{-(\epsilon_k a+b_k)}
\end{equation}
where $a$ is the Poisson parameter of the signal. 
For a single decay channel this is identical to the Bayesian prescription 
given by the Particle Data Group~\cite{pdg}. For the background estimation and
the selection efficiency, statistical and systematic errors are taken into
account by folding Gaussian distributions 
into the integration of the Poisson law used to determine the limit.

\begin{table}[hhh]
\begin{center}

\begin{tabular}{|c|c|c|c|}
\hline
 Channel & selection criteria & events  & background \\
\hline
\hline
$q^*\rightarrow q \gamma$ & 1 photon + 1 jet & 35 & 36 $\pm$ 5 \\
$q^*\rightarrow q Z_{\rightarrow e e} $ & 2 em clusters + 1 jet & 0 & 0.65 $\pm$ 0.53 \\
$q^*\rightarrow q Z_{\rightarrow \mu \mu} $ & 2 muons + 1 jet & 0 & 0.35 $\pm$ 0.05 \\
$q^*\rightarrow q Z_{\rightarrow q \bar{q}} $ & 3 jets & 32 &  25.3 $\pm$ 9.1 \\
$q^*\rightarrow q W_{\rightarrow e \nu} $ & 1 electron + 1 jet + $P_{T}^{miss}$ & 1 & 1.10 $\pm$ 0.35\\
$q^*\rightarrow q W_{\rightarrow \mu \nu} $ & 1 muon + 1 jet & 3 & 0.41 $\pm$ 0.03 \\
$q^*\rightarrow q W_{\rightarrow q \bar{q}} $ & 3 jets & 39 & 45.3 $\pm$ 17.3 \\
\hline
\end{tabular}
\caption{Number of events observed in the various $q^*$ decay channels and the corresponding 
         Standard Model expectation and total uncertainty on the mean expectation. 
         It should be noted that these numbers correspond to different invariant mass intervals, 
         as the effective mass threshold depends on the channel.}
\label{tab:res2}

\end{center}
\end{table}

Because the branching ratios of the hadronic decay of the $W$ or $Z$ bosons are dominant, the limit for the
${\tt f^*}$ decaying into a fermion and a $W$ or a $Z$ mainly depends on final states 
with at least two jets. 
So the error is dominated by the uncertainty on the absolute calibration of the calorimeters
for hadronic clusters and the uncertainty on the expected 2-jet and 3-jet cross-sections.

\begin{figure}[hhh]
\begin{center}
 \begin{tabular}{p{0.4\textwidth}p{1.0\textwidth}}
    \vspace*{-7.5cm}
      \caption[]{ \label{fig:limls1}
         {Upper limits at $95 \%$ Confidence Level on the product of the production 
          cross-section $\sigma$ and the decay branching fraction BR for excited electron $e^*$ in 
	  the various electroweak decay channels, $e \gamma$ (full line), $e Z$ (dashed line) 
	  and $\nu W$ (dotted-dashed line) as function of the excited electron mass.
          The different decay channels of the $W$ and $Z$ gauge bosons are combined. 
          Areas above the curves are excluded.}}
      &
      \mbox{\epsfxsize=0.55\textwidth
       \epsffile{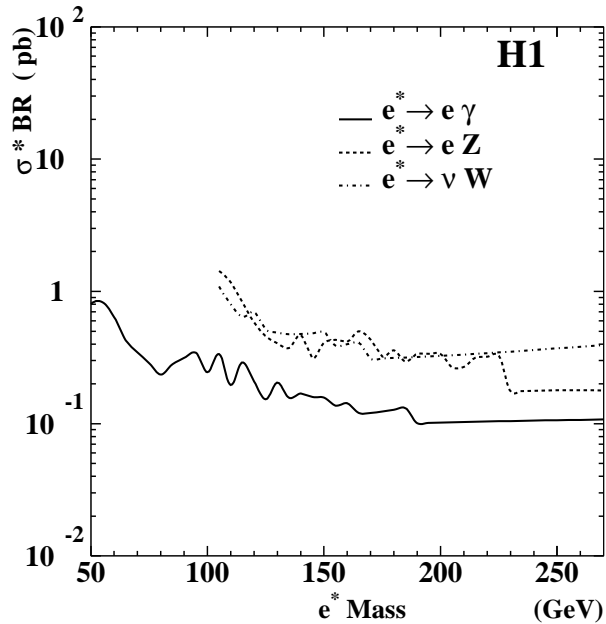}}
\end{tabular}
\end{center}
\end{figure}

\begin{figure}[hhh]
\begin{center}
 \begin{tabular}{p{0.4\textwidth}p{1.0\textwidth}}
    \vspace*{-7.5cm}
      \caption[]{ \label{fig:limls2}
          {Upper limits at $95 \%$ Confidence Level on the product of the production 
           cross-section $\sigma$ and the decay branching fraction BR for excited neutrino 
           $\nu^*$ in the various electroweak decay channels, $\nu \gamma$ (full line), 
           $\nu Z$ (dashed line) and $e W$ (dotted-dashed line) as function of the excited 
	   neutrino mass. The different decay channels of the $W$ and $Z$ gauge bosons are 
	   combined. Areas above the curves are excluded.}}
      &
      \mbox{\epsfxsize=0.55\textwidth
       \epsffile{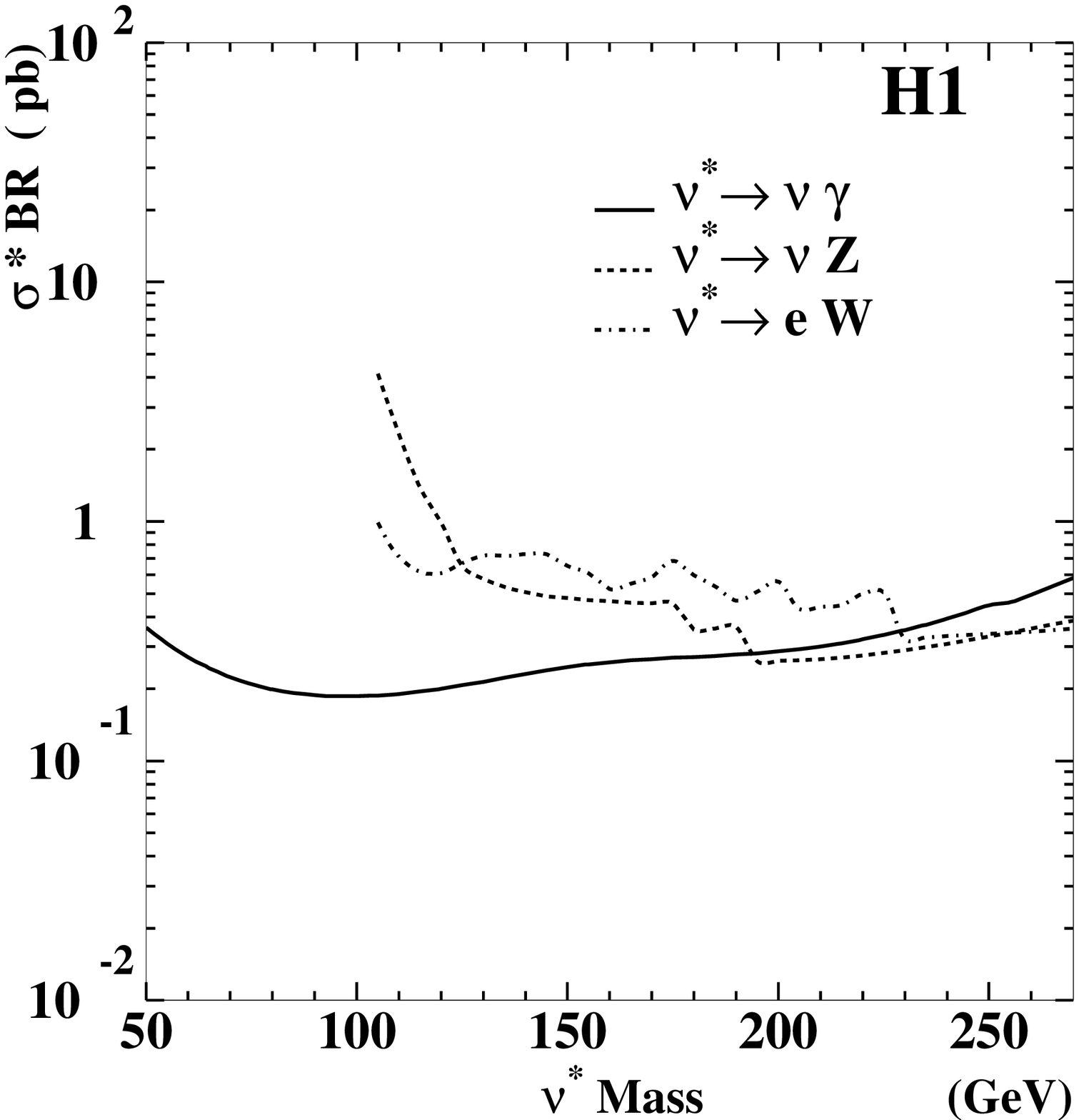}}
\end{tabular}
\end{center}
\end{figure}

\begin{figure}[hhh]
\begin{center}
 \begin{tabular}{p{0.4\textwidth}p{1.0\textwidth}}
    \vspace*{-7.5cm}
      \caption[]{ \label{fig:limqs}
          {Upper limits at $95 \%$ Confidence Level on the product of the cross-section $\sigma$ 
           and the electroweak decay channel branching ratio BR for excited quark production in 
	   the electromagnetic (full line), the $Z$ (dashed line) and $W$ (dotted-dashed line) 
	   decay channels as function of the excited quark mass. The different decay channels of 
	   the $W$ and $Z$ gauge bosons are combined. Areas above the curves are excluded.}}
      &
      \mbox{\epsfxsize=0.55\textwidth
       \epsffile{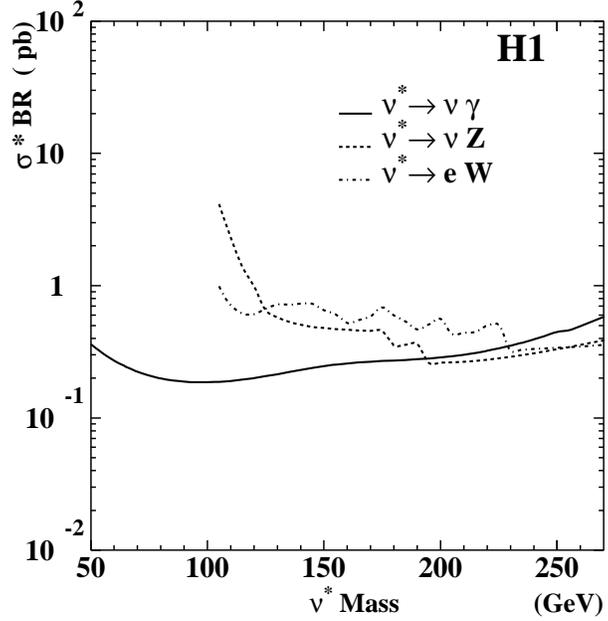}}
\end{tabular}
\end{center}
\end{figure}

The limits on the product of the ${\tt f}^*$ production cross-section and the decay branching 
fraction are shown in fig.~\ref{fig:limls1}, \ref{fig:limls2} and \ref{fig:limqs}. 
In all three cases the lowest limits are obtained using the electromagnetic decay channels. 
For the derivation of these limits it is assumed that the natural width of the ${\tt f}^*$
is much smaller than the experimental mass resolution.
For masses above 120~${\rm GeV}$, the values of these limits are below 0.2 ${\rm pb}$ for 
the $e^*$, 0.5 ${\rm pb}$ for the $\nu^*$ and 0.4 ${\rm pb}$ for the $q^*$ productions.
These results improve by an order of magnitude earlier H1 results~\cite{oldfs3}.



\boldmath\subsection{ Upper Limits on Coupling Parameters}
\label{subsec:fol}\unboldmath 

Assuming fixed numerical relations between $f$ and $f'$, the cross-section depends only 
on $f / \Lambda$ and $M_{\tt f^*}$, and thus constraints on $f / \Lambda$ can be derived.
Conventional assumptions are $f=f'$ or $f=-f'$. 
From the coupling constant relations (see equation 2) it can be seen that the coupling 
of the $e^*$ to the $ e \gamma$ mode is proportional to ($f+f'$), and that of the $\nu^*$ to 
the $\nu \gamma$ decay channel is proportional to ($f-f'$).

\begin{figure}[hhh]
\begin{center}
 \begin{tabular}{p{0.4\textwidth}p{1.1\textwidth}}
    \vspace*{-8.5cm}
      \caption[]{ \label{fig:folls1}
          {Exclusion limits on coupling constants at $95 \%$ Confidence Level as a function of the 
	   mass of the excited electron. The assumption $f = f'$ is made for the different 
	   decay channels (full, dashed and dotted-dashed lines) and for all decay channels 
	   combined (dotted line). Values of the couplings above the curves are excluded. 
           The light grey area corresponds to the exclusion domain obtained by H1 in this
           analysis. The dark grey area is excluded by the L3 experiment~\cite{lep1}.}}
      &
  \hspace*{0.5cm}   \mbox{\epsfxsize=0.55\textwidth
       \epsffile{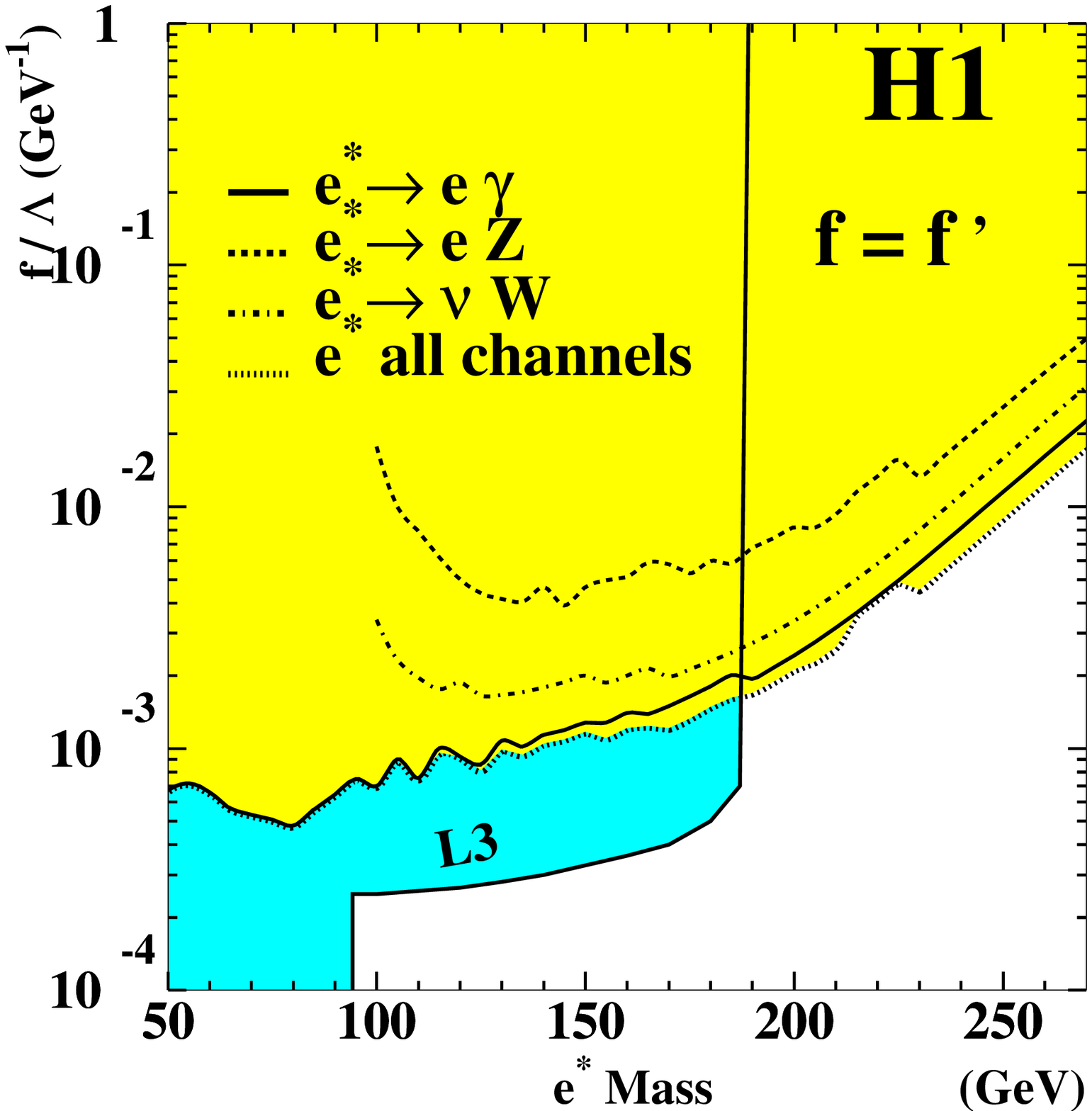}}
\end{tabular}
\end{center}
\end{figure}

In fig.~\ref{fig:folls1}, limits on the ratio $f / \Lambda$ are given for the $e^*$ for the 
hypothesis $f = f'$. We do not consider the case $f = -f'$, because the coupling constant 
$c_{\gamma e^* e}$ would be equal to 0 and the production cross-section of the $e^*$ is then 
very small.
Upper limits for $f / \Lambda$ ranging from $7 \times 10^{-4}$ ${\rm GeV}$$^{-1}$ to 
$10^{-2}$ ${\rm GeV}$$^{-1}$ are obtained at 95 $\%$ CL for an $e^*$ mass ranging from 
50 ${\rm GeV}$ to 250 ${\rm GeV}$.

\begin{figure}[hhh]
\begin{center}
\vspace*{-2cm}
\hspace*{-0.18cm}
\begin{tabular}{cc}
 
\epsfxsize=0.52\textwidth
 \epsffile{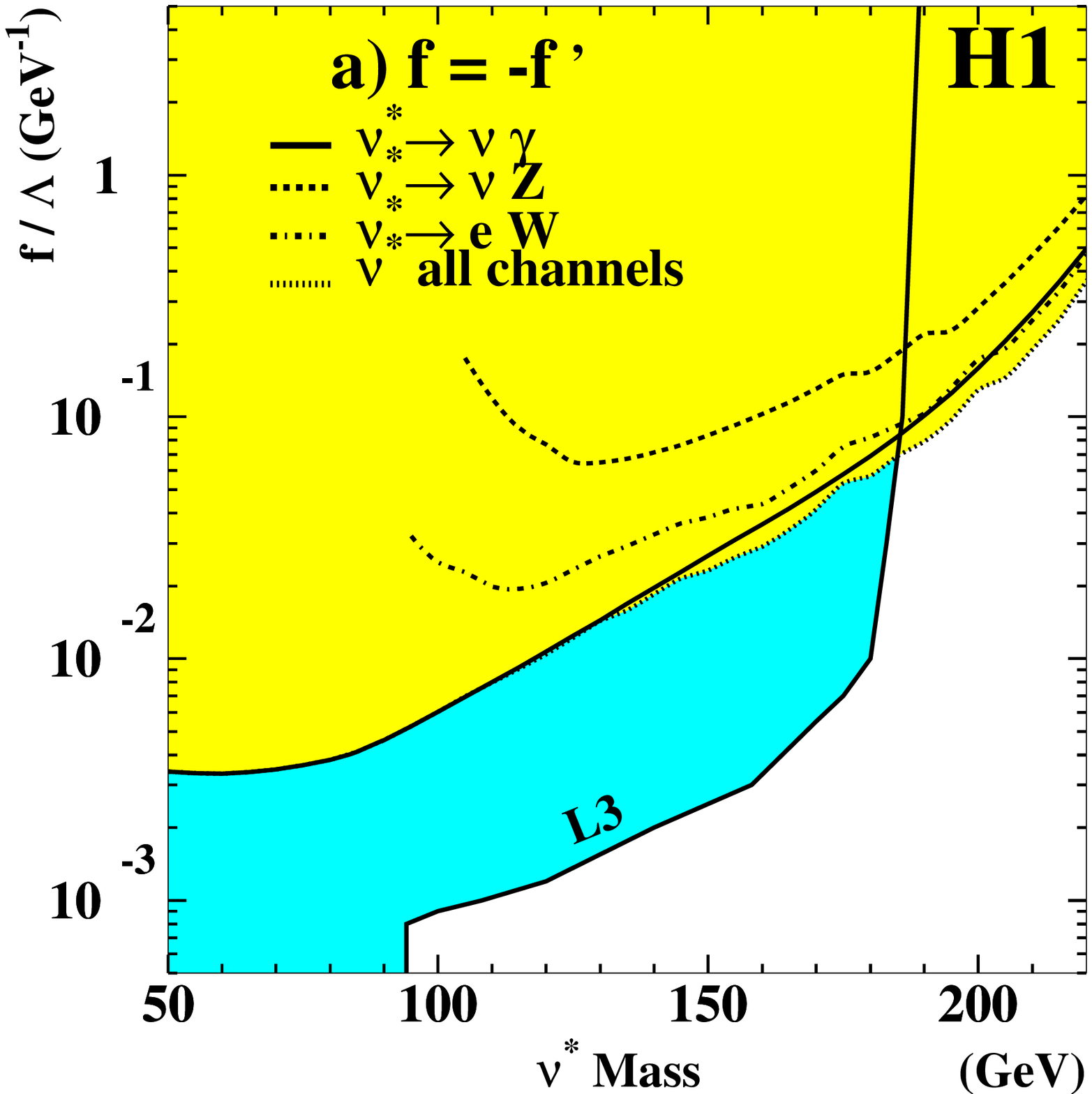} &
\epsfxsize=0.52\textwidth

\hspace*{-0.3cm}\epsffile{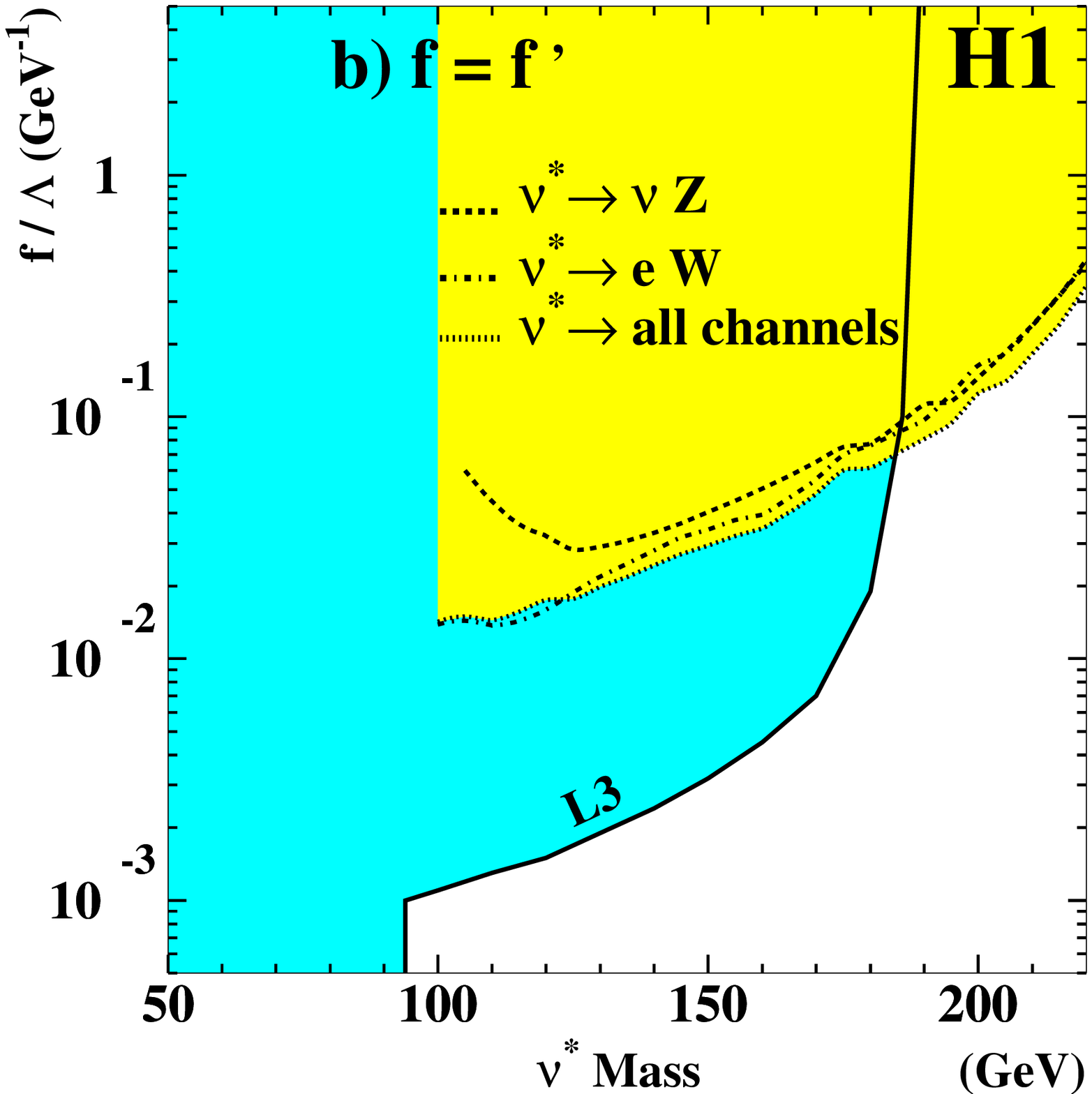} \\
\end{tabular}

\caption[]{Exclusion limits on coupling constants at $95 \%$ Confidence Level
as a function of the mass
of the excited neutrino. 
The assumptions $f = - f'$ and $f =  f'$ are made for figures ({\it a}) and ({\it b}) respectively. 
The results for the different decay channels are shown separately (full, dashed and dotted-dashed lines) 
and for all decay channels combined (dotted line).
Values of the couplings above the curves are excluded. 
The light grey area corresponds to the exclusion domain obtained by H1 in this analysis.
The dark grey area is excluded by the L3~\cite{lep1} experiment.}

\label{fig:folls2}
\end{center}
\end{figure}

In fig.~\ref{fig:folls2}, limits on the ratio $f / \Lambda$ are given for $\nu^*$, assuming  
$f = -f'$ and $f = f'$. These two assumptions correspond to very different $\nu^*$ branching
ratios, as shown in table~\ref{tab:br} for the example of two $\nu^*$ masses. In particular 
when $f = f'$ the $\nu^* \rightarrow e \gamma$ has a branching ratio equal to 0.
Somewhat better limits are obtained when $f = -f'$. 
The values of the limits for $f / \Lambda$ vary between 
$3 \times  10^{-3}$ to $ 10^{-1}$ ${\rm GeV}$$^{-1}$ for an $\nu^*$ mass ranging from 
50 to 200 ${\rm GeV}$. These upper limits are conservative for masses above $\simeq 170 \GeV$
where the narrow-width approximation underestimates the total $\nu^*$ production
cross-section.

Assuming $f / \Lambda=  1 / M_{l^*}$, masses below 223 and 114 ${\rm GeV}$ are excluded
at 95 $\%$ CL for the $e^*$ ($f = f'$) and $\nu^*$ ($f = -f'$)
production, respectively.

\begin{table*}[hhh]
  \renewcommand{\doublerulesep}{0.4pt}
  \renewcommand{\arraystretch}{1.2}
 \begin{center}
 \begin{tabular}{p{0.45\textwidth}p{0.55\textwidth}}

    \caption
    {\label{tab:br}
    Branching ratios in $\%$ of the $\nu^*$ decay modes for different relations 
    between $f$ and $f'$.} &

\begin{tabular}{|c|c|c|}
\hline
$M_{\nu^*}$ $({\rm GeV})$ & 100 & 200  \\
\hline
\multicolumn{3}{|c|}{ $f = f'$} \\	
\hline
$\nu^*\rightarrow \nu \gamma$ & 0 & 0  \\
$\nu^*\rightarrow \nu Z$ & 13 & 37 \\
$\nu^*\rightarrow e W$ & 87 & 63\\
\hline
\multicolumn{3}{|c|}{ $f = - f'$} \\
\hline
$\nu^*\rightarrow \nu \gamma$ &72 &34  \\
$\nu^*\rightarrow \nu Z$ & 1 & 10 \\
$\nu^*\rightarrow e W$ & 27 & 56 \\
\hline
\end{tabular}
  \end{tabular}
  \end{center}
\end{table*}

\begin{figure}[hhh]
\begin{center}
 \begin{tabular}{p{0.4\textwidth}p{1.0\textwidth}}
    \vspace*{-8.5cm}
      \caption[]{ \label{fig:folqs}
          {Exclusion limits on coupling constants at $95 \%$ Confidence Level
as a function of the mass of the excited quark, assuming $f = f'$ and $f_s = 0$.
The results for the different electroweak decay channels are shown separately (full, dashed line 
and dotted-dashed lines) and for all decay channels combined (dotted line).
Values of the couplings above the curves are excluded. The light grey area corresponds
to the exclusion domain obtained by the H1 experiment in this analysis. 
The dark grey area is excluded by the DELPHI experiment~\cite{lep2} 
assuming that the branching ratio of the $q^* \rightarrow q \gamma$ is equal to 1. }}
      &
      \mbox{\epsfxsize=0.55\textwidth
       \epsffile{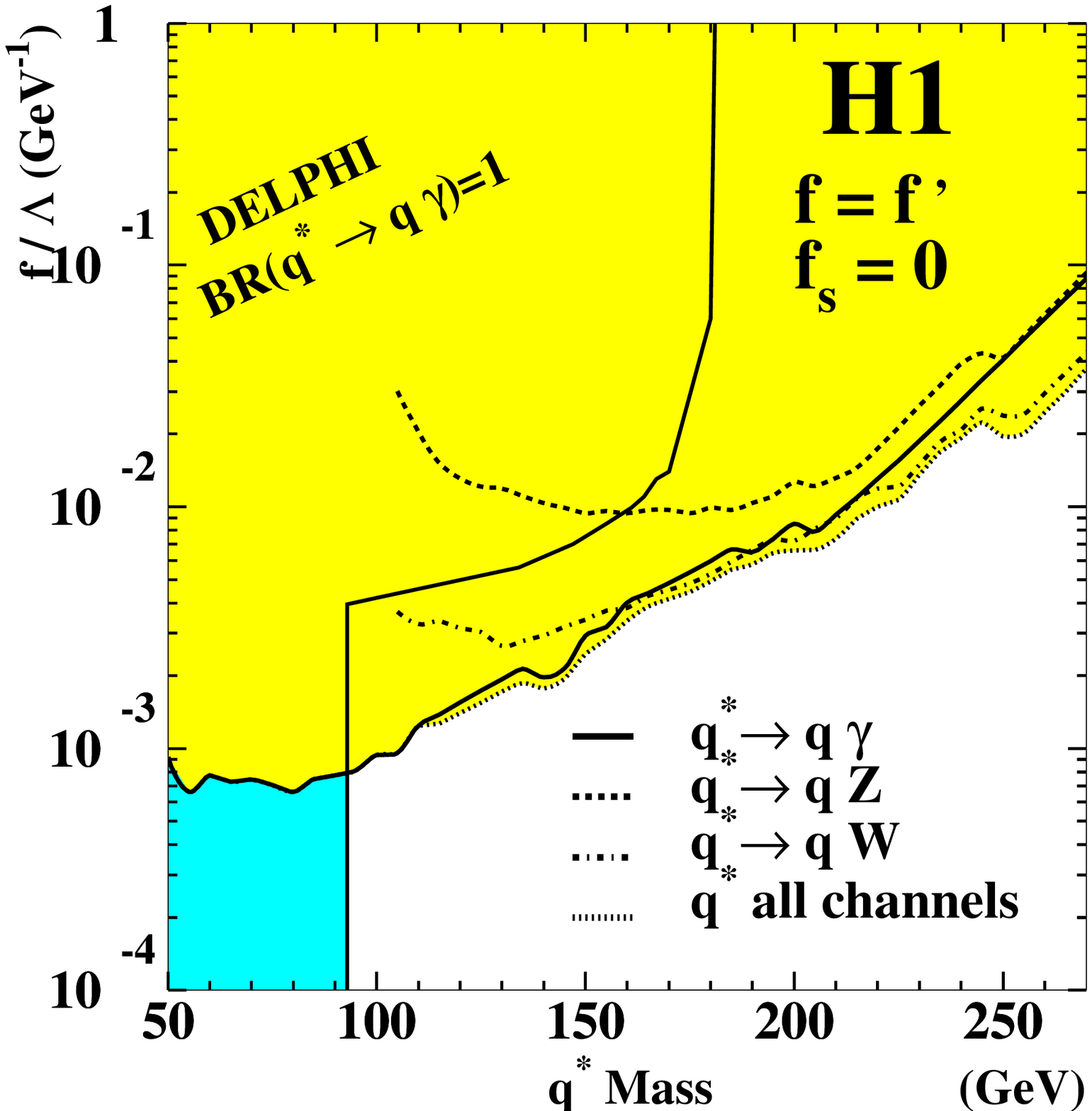}}
\end{tabular}
\end{center}
\end{figure}

\begin{figure}[hhh]
\begin{center}
 \begin{tabular}{p{0.4\textwidth}p{1.0\textwidth}}
    \vspace*{-7.5cm}
      \caption[]{ \label{fig:valfs}
          {Exclusion limits on $f$ values at $95 \%$ Confidence Level
           as a function of the mass of the excited quark, assuming $\Lambda=M(q^*)$, $f = f'$ 
	   and for different $f_s$ values. Exclusion limits from CDF (the 2 right curves) for 
	   2 $f_s$ values have been derived from table 1 of reference~\cite{cdf2}. 
           Values of the couplings above the curves are excluded.}}
      &
      \mbox{\epsfxsize=0.55\textwidth
       \epsffile{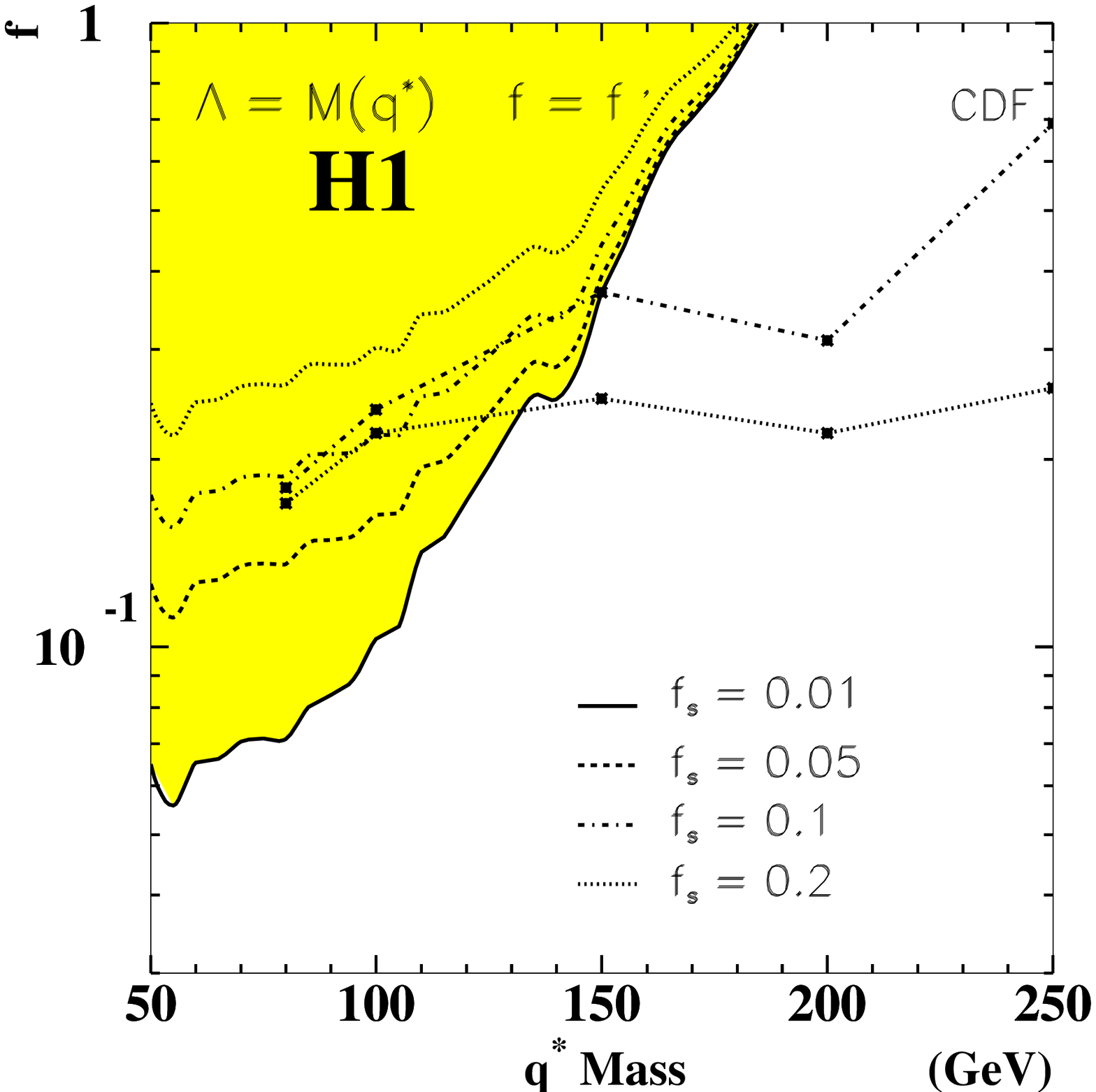}}
\end{tabular}
\end{center}
\end{figure}

Limits for the $q^*$ assuming $f = f'$
and only electroweak couplings (i.e. $f_s = 0$) are shown in fig.~\ref{fig:folqs}.
The exclusion limits for $q^*$ masses between 50 to 250 ${\rm GeV}$ corresponds to values of
$f / \Lambda$ between $9 \times  10^{-4}$ and $ 2 \times  10^{-2}$ ${\rm GeV}$$^{-1}$. 
Assuming $f / \Lambda=  1 / M_{q^*}$, masses below 188 ${\rm GeV}$ are excluded
at 95 $\%$ CL.
The $f_s = 0$ assumption allows to make a study which is complementary to the analysis done by the 
CDF experiment~\cite{cdf1,cdf2} at the Tevatron.
At a $p\overline{p}$ collider excited quarks are produced in a quark-gluon fusion mechanism
which requires $f_s \neq$ 0. Assuming $\Lambda = M_{q^*}$, CDF excludes excited quarks
in the mass range 80-300 ${\rm GeV}$ for $f = f' = f_s$ values greater than 0.2 and up to 760 ${\rm GeV}$ 
if $f = f' = f_s = 1 $.\\
The complementarity of the H1 results to those of CDF is illustrated
on fig.~\ref{fig:valfs}, where a comparison of the exclusion domains 
of $f=f'$ values, obtained by CDF and H1, is shown for different hypotheses 
on the $f_s$ value. As soon as $f_s$ is smaller than $\sim 0.1$ and for 
$M_{q^*} \leq$ 130 ${\rm GeV}$, our analysis probes a domain not excluded by 
Tevatron experiments. 


LEP experiments have also reported searches for excited leptons at $e^+e^-$
center of mass energies up to $\sqrt{s}=$ 189 ${\rm GeV}$~\cite{lep1,lep2,lep3}.
The results for excited leptons produced in pairs, when the limits 
are independent of $f$ and $f'$, as well as results for single $e^*$ and $\nu^*$
production assuming $f=f'$ or $f=-f'$ are included in fig.~\ref{fig:folls1} and \ref{fig:folls2}. 
In~\cite{lep3} results independent from a hypothesis on the relation between $f$ and $f'$ are also given.
Fig.~\ref{fig:folqs} shows the result of a $q^*$ search~\cite{lep2} at LEP,
assuming a branching ratio of 1 for $q^* \rightarrow q \gamma$. 

\boldmath\section{ Summary}\label{subsec:sum} \unboldmath
Using $e^+p$ data taken from 1994 to 1997 corresponding to an integrated luminosity of 37 pb$^{-1}$,
a search for production of excited fermions has been carried out.
No evidence has been found for excited electrons, excited neutrinos or excited quarks 
for decays into any of the gauge bosons $\gamma$, $Z$, $W$ and Standard Model fermions.

New limits for the production of excited fermions have been obtained, which
improve previous H1 results by a factor 10 and previous published ZEUS 
results~\cite{zeus1} based on 4 times smaller integrated luminosity. 
For masses above 180 ${\rm GeV}$, i.e. in a domain extending beyond the kinematic reach of LEP, 
compositeness scales of $f / \Lambda$ in the range of 
$1.5 \times 10^{-3}$ to $2 \times 10^{-2}$ ${\rm GeV}$$^{-1}$
are excluded from the search for excited electrons, and $5.6 \times 10^{-2}$ to 
0.32 ${\rm GeV}$$^{-1}$ for excited neutrinos. 
Assuming $f / \Lambda=  1 / M_{\tt f^*}$, excited fermions with masses below 223, 114, and 188 ${\rm GeV}$, 
for $e^*$, $\nu^*$ and $q^*$ productions, respectively, are excluded.
The results obtained on $q^*$ production via electroweak couplings are 
complementary to the results obtained at the Tevatron $p \bar{p}$ collider 
where $q^*$ production via strong coupling is investigated.

\vspace*{1cm}

\section*{Acknowledgements}
We are grateful to the HERA machine group whose outstanding efforts
made this experiment possible.
We appreciate the immense effort of the engineers and technicians who
constructed and maintain the detector.
We thank the funding agencies for their financial support of the experiment.
We wish to thank the DESY directorate for the hospitality extended
to the non-DESY members of the Collaboration.



\vfill
\clearpage

\begin{thebibliography}{99}

\bibitem{co1}S. Weinberg, \Journal{\PRD}{13}{1976}{974};
${\it ibid}$ \Journal{\PRD}{19}{1979}{1277};\\
L. Susskind, \Journal{\PRD}{20}{1979}{2619};\\
E. Farhi and L. Susskind, \Journal{\PRD}{20}{1979}{3404};\\
H. Harrari, \Journal{\PLB}{86}{1979}{83};\\
H. Harrari and N. Seiberg, \Journal{\PLB}{98}{1981}{269};\\
H. Fritzsch and G. Mandelbaum, \Journal{\PLB}{102}{1981}{319};
${\it ibid}$, \Journal{\PLB}{109}{1982}{224}.

\bibitem{oldfs1}H1 Collaboration, I. Abt et al., \Journal{\NPB}{396}{1993}{3};\\
H1 Collaboration, T. Ahmed et al., \Journal{\PLB}{340}{1994}{205}.

\bibitem{oldfs3}H1 Collaboration, S. Aid et al., \Journal{\NPB}{483}{1997}{44}.

\bibitem{zeus1}ZEUS Collaboration, J. Breitweg et al., \Journal{\ZPC}{76}{1997}{631}.

\bibitem{kuhn}J. K\"{u}hn and P. Zerwas, \Journal{\PLB}{147}{1984}{189}.

\bibitem{jikia}G. Jikia, \Journal{\NPB}{333}{1990}{317}.

\bibitem{hagi}K. Hagiwara, S. Komamiya and D. Zeppenfeld, \Journal{\ZPC}{29}{1985}{115}.

\bibitem{baur}U. Baur, M. Spira and P. Zerwas, \Journal{\PRD}{42}{1990}{815}.

\bibitem{boud}F. Boudjema, A.Djouadi, J.L. Kneur,
\Journal{\ZPC}{57}{1993}{425}.

\bibitem{brod}S.J. Brodsky and S.D. Drell, \Journal{\PRD}{22}{1980}{2236}.

\bibitem{rena}F.M Renard, \Journal{\PLB}{116}{1982}{264}. 

\bibitem{dete} H1 Collaboration, I. Abt et al., 
\Journal{\NIMA}{386}{1997}{310}, ${\it ibid}$, 348.

\bibitem{lar} H1 Calorimeter Group, B. Andrieu et al., \Journal{\NIMA}{336}{1993}{460}.

\bibitem{testcalo1} H1 Calorimeter Group, B. Andrieu et al., \Journal{\NIMA}{344}{1994}{492}.

\bibitem{testcalo2} H1 Calorimeter Group, B. Andrieu et al., \Journal{\NIMA}{350}{1994}{57};
${\it ibid}$, \Journal{\NIMA}{336}{1993}{499}.

\bibitem{spacal} H1 SPACAL Group, R.D. Appuhn et al., \Journal{\NIMA}{386}{1997}{397}.

\bibitem{bemc} H1 BEMC Group, J. Ban et al., \Journal{\NIMA}{372}{1996}{399}.

\bibitem{iron} W. Braunschweig et al., \Journal{\NIMA}{270}{1988}{334}.

\bibitem{djan}DJANGO 6.2: G.A. Schuler and H. Spiesberger, Proc. of the Workshop
``Physics at HERA'' (Eds. W. Buchm\"{u}ller and G.Ingelman),
DESY Hamburg 1991,Vol 3, p 1419. 

\bibitem{hera}HERACLES 4.4: A. Kwiatkowski, H. Spiesberger and H.-J. M\"{o}ring,
\Journal{\CPC}{69}{1992}{155}.

\bibitem{aria} ARIADNE 4.0: L. L\"{o}nnblad, \Journal{\CPC}{71}{1992}{15}.

\bibitem{cdm} B. Andersson, G. Gustafson, L. L\"{o}nnblad,
\Journal{\ZPC}{43}{1989}{625}.

\bibitem{rapg} H. Jung, \Journal{\CPC}{86}{1995}{147}, Version 2.06,
http://www-h1.desy.de/$\sim$jung/rapgap.html.

\bibitem{mrst} A.D. Martin, R.G. Roberts, W.J. Stirling and R.S. Thorne,
\Journal{\EJC}{4}{1998}{463}.

\bibitem{mrst1} H1 Collaboration, S. Aid et al.,
\Journal{\NPB}{470}{1996}{3}.

\bibitem{mrst2} ZEUS Collaboration, M. Derrick et al.,
\Journal{\ZPC}{72}{1996}{399}.


\bibitem{jets} JETSET 7.3 and 7.4: T. Sj\"{o}strand, Lund Univ. preprint LU-TP-95-20
 (August 1995) 321pp; {\it ibid}, CERN preprint TH-7112-93 (February 1994) 305pp.


\bibitem{comp1} EPCOMPT: F. Raupach, Proc. of the Workshop ``Physics at HERA''
                (Eds. W. Buchm\"{u}ller and G. Ingelman), DESY Hamburg 1991, Vol 3, p 1473.

\bibitem{comp2} A new Generator for Wide Angle Bremsstrahlung: Ch. Berger and P. Kandel, 
Proc. of the Monte Carlo Generators for HERA Physics Workshop, DESY-PROC-1999-02, p 596.

\bibitem{pyth} PYTHIA 5.7: T. Sj\"{o}strand, CERN-TH-6488 (1992), \Journal{\CPC}{82}{1994}{74}.

\bibitem{lpair} LPAIR: S. P Baranov et al., Proc. of the Workshop ``Physics at HERA'',
 (Eds. W. Buchm\"{u}ller and G.Ingelman), DESY Hamburg 1991,Vol 3, p 1478.

\bibitem{dirk} GRAPE-Dilepton, A generator for Dilepton Production in ep collisions, 
               T. Abe et al, Proc. of the Workshop ``Monte Carlo Generators for HERA Physics''
               (Eds. A.T.~Doyle, G.~Grindhammer, G.~Ingelman and H.~Jung), 
               DESY-PROC-1999-02, p 566; 
               Lepton Pair Monte Carlo Generators for HERA Physics, 
               D. Hoffmann, L. Favart, {\it ibid} p 575.
 
\bibitem{epvec} EPVEC: U. Baur, J.A.M Vermaseren, D. Zeppenfeld, \Journal{\NPB}{375}{1992}{3}.

\bibitem{kohl} T.~K\"{o}hler, Proc. of the Workshop ``Physics at HERA''
               (Eds. W. Buchm\"{u}ller and G.Ingelman), DESY Hamburg 1991, Vol 3, p 1526.

\bibitem{isa} I. N\'egri, "Recherche de fermions excit\'es dans l'exp\'erience H1 aupr\`es du
collisionneur positron-proton HERA", Th\`ese de Doctorat, Universit\'e de la M\'editerran\'ee, April 1998.

\bibitem{muon} H1 Collaboration, C. Adloff et al., \Journal{\EJC}{5}{1998}{575}.

\bibitem{hrap} H1 Collaboration, C. Adloff et al.,
\Journal{\EJC}{13}{2000}{415}.

\bibitem{work} Proc. of the Workshop ``Monte-Carlo Generators for HERA Physics''
               (Eds. A.T.~Doyle, G.~Grindhammer, G.~Ingelman and H.~Jung), 
               QCD Cascade Working Group, DESY-PROC-99-02.

\bibitem{pdg}R.M.~Barnett et al., \Journal{\PRD}{54}{1996}{1};
O.~Helene, \Journal{\NIMA}{212}{1983}{319}.

\bibitem{hipt} H1 Collaboration, C. Adloff et al.,
\Journal{\ZPC}{74}{1997}{191}.

\bibitem{errjets} T. Carli, 
                  ``Renormalisation scale dependencies in di-jet production at HERA'', 
                  Proc. of the Workshop ``Monte-Carlo Generators for HERA Physics''
		  (Eds. A.T.~Doyle, G.~Grindhammer, G.~Ingelman and H.~Jung), 
                  DESY-PROC-99-02, p 184;
                  \, P. Bate, 
		  ``High Transverse Momentum 2-jet and 3-jet Cross section Measurements 
                  in Photoproduction'',
                  Ph.D. thesis, The University of Manchester, December 1999;
                  \, H1 Collaboration, 
		  ``First Measurement of Three jet Cross Sections in DIS at HERA'', 
                  Contrib. paper \#157 to the International Europhysics  Conference on High
                  Energy Physics, Tampere, Finland (15-21 July 1999) 12pp.

\bibitem{hel92} T.~Helbig and H.~Spiesberger, 
\Journal{\NPB}{373}{1992}{73}.

\bibitem{cdf1}CDF Collaboration, F.~Abe et al., 
\Journal{\PRD}{55}{1997}{5263}.

\bibitem{cdf2}CDF Collaboration, F.~Abe et al., 
\Journal{\PRL}{72}{1994}{3004}.

\bibitem{lep1}L3 Collaboration,  M.~Acciarri et al.,
\Journal{\PLB}{473}{2000}{177}.

\bibitem{lep2}DELPHI Collaboration, P.~Abreu et al., \Journal{\EJC}{8}{1999}{41}.

\bibitem{lep3} OPAL Collaboration, G.~Abbiendi et al., 
\Journal{\EJC}{14}{2000}{73}.

\end{thebibliography}
\end{document}